\documentclass[acmsmall]{acmart}

\usepackage{svg}
\usepackage{makecell}
\usepackage{graphicx}
\usepackage{subcaption}
\usepackage{enumitem}

\usepackage[frozencache,cachedir=minted-cache]{minted} 

\usepackage{xcolor}
\usepackage{multirow}
\usepackage[most]{tcolorbox}

\definecolor{brainstorm}{HTML}{FFB347}
\definecolor{summarise}{HTML}{966FD6}
\definecolor{elaborate}{HTML}{71C562}
\definecolor{draft}{HTML}{FB6B89}
\definecolor{freewrite}{HTML}{B39EB5}
\definecolor{associate}{HTML}{D64A4A}
\definecolor{custom}{HTML}{B7AFA3}
\definecolor{custom2}{HTML}{E8D0A9}

\definecolor{keyword}{HTML}{CED8DF}
\definecolor{concept}{HTML}{FFB5B7}
\definecolor{sticky_note}{HTML}{748B97}
\definecolor{section}{HTML}{0099FF}
\newcommand{\revision}[1]{{\color{black}#1}}
\newcommand{\minor}[1]{{\color{black}#1}}

\newcommand{\RNum}[1]{\uppercase\expandafter{\romannumeral #1\relax}}
\AtBeginDocument{%
  }

\setcopyright{acmlicensed}
\acmJournal{PACMHCI}
\acmYear{2025} \acmVolume{9} \acmNumber{7} \acmArticle{316} \acmMonth{11}\acmDOI{10.1145/3757497}
\begin{document}

\title{Polymind: Parallel Visual Diagramming with Large Language Models to Support Prewriting Through Microtasks}

\author{Qian Wan}
\orcid{0000-0002-4250-8780}
\affiliation{%
  \institution{City University of Hong Kong}
  \streetaddress{83 Tat Chee Avenue}
  \city{Hong Kong}
  \country{Hong Kong}
}
\email{qianwan3-c@my.cityu.edu.hk}

\author{Jiannan Li}
\orcid{https://orcid.org/0000-0001-8409-4910}
\affiliation{%
  \institution{Singapore Management University}
  \city{Singapore}
  \country{Singapore}
  }
\email{jiannanli@smu.edu.sg}

\author{Huanchen Wang}
\orcid{0000-0001-9339-1941}
\affiliation{
  \department{Department of Computer Science}
  \institution{City University of Hong Kong}
  \city{Hong Kong}
  \country{China}
}

\affiliation{
  \department{Department of Computer Science and Engineering}
  \institution{Southern University of Science and Technology}
  \city{Shenzhen}
  \state{Guangdong}
  \country{China}
}
\email{hc.wang@my.cityu.edu.hk}

\author{Zhicong Lu}
\orcid{https://orcid.org/0000-0001-8409-4910}
\affiliation{%
  \institution{George Mason University}
  \city{Fairfax}
  \country{USA}
}
\email{zhiconlu@cityu.edu.hk}

\renewcommand{\shortauthors}{Qian Wan, Jiannan Li, Huanchen Wang, and Zhicong Lu} 


\begin{abstract}
Prewriting is the process of generating and organising ideas before a first draft. It consists of a combination of informal, iterative, and semi-structured strategies such as visual diagramming, which poses a challenge for collaborating with large language models (LLMs) in a turn-taking conversational manner. We present \textit{Polymind}, a visual diagramming tool that leverages multiple LLM-powered agents to support prewriting. The system features a parallel collaboration workflow in place of the turn-taking conversational interactions. It defines multiple ``microtasks'' to simulate group collaboration scenarios such as collaborative writing and group brainstorming. Instead of repetitively prompting a chatbot for various purposes, \textit{Polymind} enables users to orchestrate multiple microtasks simultaneously. Users can configure and delegate customised microtasks, and manage their microtasks by specifying task requirements and toggling visibility and initiative. Our evaluation revealed that, compared to ChatGPT, users had more customizability over collaboration with \textit{Polymind}, and were thus able to quickly expand personalised writing ideas during prewriting.
\end{abstract}



\received{July 2024}
\received[revised]{December 2024}
\received[accepted]{March 2025}


\begin{teaserfigure}
    \centering
    \includegraphics[width=\linewidth]{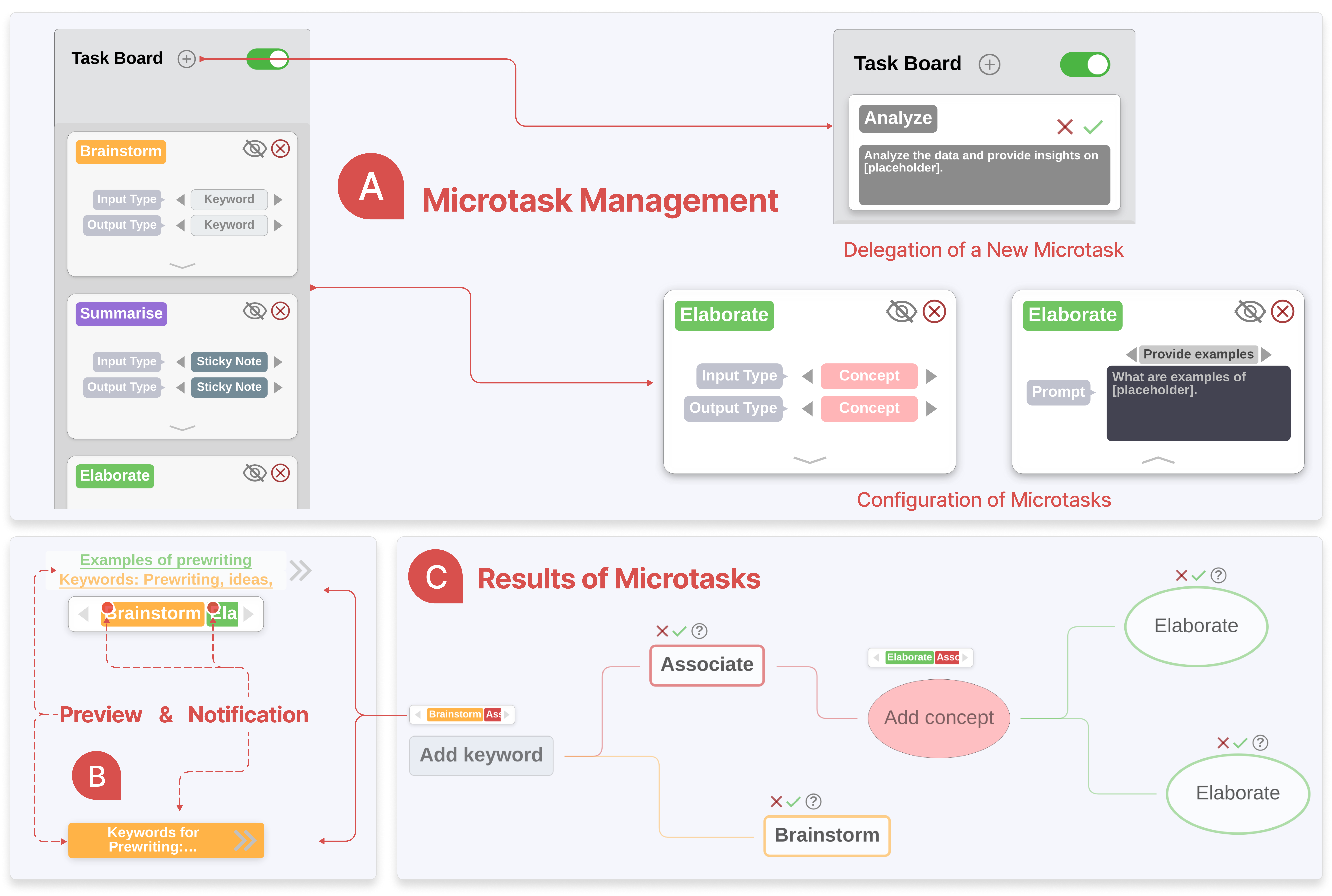}
    \caption{The microtasking workflow of \textit{Polymind}: A) A user can delegate new microtasks, and configure microtasks by specifying input \& output types, prompts, initiative modes, etc. B) An active microtask notifies users when results are ready and provides previews on the canvas. C) Once expanded, resulting diagrams are displayed using a hollow shape in contrast to user created diagrams, and distinctive border colours indicating which microtasks they are from.}
    \label{fig:teaser}
\end{teaserfigure}

\maketitle

\section{Introduction}
In creative ideation, people often apply semi-structured diagrammatic methods, such as mind mapping, concept mapping, and argument mapping~\cite{baroudy2008procedural,mogahed2013planning}.
These thinking strategies support humans' visual exploration of key topics, concepts, arguments, and their relationships~\cite{lu2018inkplanner} by connecting, arranging, and grouping the corresponding graphical representations.


In addition to using visual thinking tools, creativity-related research has highlighted that certain global thinking strategies can enhance creativity.
One strategy that has been found beneficial for creative thinking in numerous domains~\cite{lo2017critical,dell2020four}, such as writing~\cite{al2018effect,al2015effect} and problem solving~\cite{peterson1998parallel}, is \textit{parallel thinking}~\cite{bono1985six,de2017six}.
It encourages systematic exploration and evaluation of ideas with distinct modes of thought.
For ideation activities, two modes of thinking are of key significance: divergent and convergent thinking~\cite{runco2023creativity}. The former involves generating multiple solutions, while the latter focuses on working out a single correct or best solution to a given problem.
However, successfully applying both convergent and divergent thinking simultaneously often requires considerable training~\cite{de2017six,gregory2012real}, as it breaks away from traditional cognitive patterns by asking for an explicit switch of mental modes.
In practice, people often rely on group collaboration to achieve convergent and divergent thinking in parallel, where individual group members offer separate angles of thought.
Previous creativity literature has further proved that real-time parallel inputs in human collaboration can produce significantly more unique ideas than serial interaction interfaces ~\cite{shih2009groupmind,hymes1992unblocking}, and presenting multiple ideas in parallel can stimulate more productive and creative outcomes than serial presentation ~\cite{dow2010parallel}.

We are thus interested in exploring  human-AI collaborative ideation where ``virtual collaborators'' powered by generative artificial intelligence enhance visual diagramming and facilitate parallel thinking.
Our effort focused on the creative act of \textit{prewriting}, the first stage of the writing process where writers discover and develop new ideas before they are ``\textit{ready for the words and the page}''~\cite{rohman1965pre}.
Research has shown prewriting strategies~\cite{barnett1989writing,rohman1965pre} can improve the quality of final documents by relieving memory strain~\cite{flower1981cognitive,kellogg1990effectiveness} and incubating new ideas with organised schema of existing knowledge~\cite{lorenz2009using,davies2011concept}.

Authors often apply visual diagramming to expand and organize their ideas in prewriting~\cite{mogahed2013planning}.
There has been a growing interest in leveraging Large Language Models (LLMs) to support writing in general~\cite{chung2022talebrush,clark2018creative,zhang2023visar,suh2023sensecape} and specifically to enhance writers' idea diversity~\cite{benharrak2024writer, shaer2024ai}.
However, conversational LLM interfaces, as exemplified by ChatGPT, fit poorly into diagramming-based ideation workflows.
A conversational interface removes its users from the often diagrammatic and divergent thinking context, making it difficult for users to leverage existing information represented in the diagrams.
Furthermore, productive parallel thinking requires systematically applying multiple distinct angles of thoughts over the ideas at hand. 
For a pure-text, turn-taking conversational interface, this entails that a user has to remember the specific prompt for each angle and repeatedly apply them throughout the often iterative prewriting process~\cite{rohman1965pre}.
Although several previous systems have already incorporated diagrammatic interfaces and multiple distinct LLM operations for prewriting (e.g., ~\cite{lu2018inkplanner,zhang2023visar,suh2023sensecape}), they are still far from achieving parallel thinking effects in that the interactions remained a serial manner where each LLM operations waited for users to switch into different modes of thoughts.
In this paper, we introduce \textit{Polymind}, a visual diagramming tool infused with multiple distinct LLM agents working in parallel to facilitate prewriting.
\textit{Polymind} supports a writer's initiative in listing, arranging, and connecting ideas via a diagramming interface while augmenting this visual thinking process in-situ with LLM-powered parallel angles of thought.
These angles cover a repertoire of key prewriting techniques involving both convergent and divergent thinking, such as ideation, association, summarisation, drafting, as informed by our literature review of both writing and creativity support tools.
We operationalize these techniques through small, manageable, and independent ``microtasks'' that LLMs execute over diagrams nodes as the writer's virtual collaborators. Microtasking is commonly used in human group collaboration, including crowdsourcing~\cite{latoza2014microtask,chen2017retool}, collaborative writing \cite{iqbal2018multitasking,birnholtz2013write,teevan2016supporting}, and group brainstorming \cite{chilton2019visiblends,teevan2016supporting}, as it proves to effectively simplify complex tasks \cite{cheng2015break,kokkalis2013taskgenies}, facilitate recovery from interruptions, and produce higher quality results \cite{cheng2015break}.
In \textit{Polymind} these microtasks process ideas and their relationships as represented in the diagrams to expand, transform, and re-organize them.
The results are reflected back in the diagrams as new or modified nodes and connections. 
Further, \textit{Polymind} allows users to run and orchestrate multiple microtasks simultaneously to achieve efficient expansion and iteration of ideas.  
\revision{Our evaluation involved 10 non-expert writers with limited technical background performing two prewriting sessions. Results showed that \textit{Polymind} was overall seen as more creative and customisable for prewriting than a turn-taking conversational interface.}
Parallel LLM assistance allowed participants to efficiently leverage both convergent and divergent modes of thinking.
This lent support to a number of new ideation capabilities, including quick expansion of existing diagrams (e.g., a concept tree).
Meanwhile, our mixed-initiative workflows granted easier and faster control over LLM generation on a prewriting canvas.
Our study results also highlighted more nuanced design considerations around parallel LLM assistance for human-AI co-creativity, suggesting the value of dynamically adjusting AI pro-activity levels and balancing awareness of AI actions and non-intrusiveness to human workflows.


Our paper makes the following contributions to HCI:
\begin{itemize}[topsep=0pt]
\item a human-AI collaborative ideation workflow with multiple, heterogeneous AI-based microtaskers to achieve parallel thinking for visual diagramming.
\item a set of user interface designs to support the management of these microtasks in a mixed-initiative manner to support users' creative agency and control.
\item a functional implementation of the workflow and  interface designs as an LLM-based, microtask-enhanced diagramming tool for creative prewriting.
\item a user study that shows the feasibility and potential of the microtasking workflow
\end{itemize}

\section{Related Work}
In this section, we briefly review related work on human-AI co-creativity, writing support, and recent large large language models augmented writing tools.

\subsection{From Human-human Co-creativity to Human-AI Co-creativity}
Supporting human co-creativity scenarios such as group brainstorming have been extensively explored in CSCW. Early systems generally used visual or cross-modality stimuli to facilitate ideation in face-to-face brainstorming sessions. For example, Wang et al. proposed IdeaExpander, which was deployed in a group chat to process group conversations, and retrieve related pictures to stimulate ideation~\cite{wang2010idea}. It later proved to produce significantly more ideas than no picture support in a follow-up study~\cite{wang2011diversity}. Shi et al. proposed IdeaWall that continuously presents semantic and visual stimuli by extracting key information from verbal discussions~\cite{shi2017ideawall}. Although no significant enhancement of diversity or quantity of ideas was reported, they found that pictorial support can significantly stimulate group dynamics, which were measured as duration and quantity of lulls during conversations.

Another line of research aims to leverage group dynamics to make brainstorming more productive. Wang et al., for example, probed the cultural differences as a source of conceptual diversity~\cite{wang2011diversity}. Their study found that intercultural brainstorming pairs can cover significantly broader concepts than intracultural pairs given proper visual stimuli. Chan et al. proposed a groupware that augmented crowd innovation with real-time expert guidance~\cite{chan2016improving}. They found that experienced expert facilitators increased both the quantity and creativity of crowd workers' ideas compared to the group without facilitators. Shih et al.~\cite{shih2009groupmind} designed GroupMind that enabled multiple users to contribute in parallel to a digital mind-mapping whiteboard. In a controlled lab study, they found that the parallel-input workflow can produce significantly more unique ideas than serial inputs.

With the advance of AI, especially generative AI models, how they might shape creative activities and influence creative output has become a question of research interest. Wan et al. studied the human-AI co-creativity process during the prewriting stage~\cite{wan2024felt}. They identified three key stages of collaboration with large language models: Ideation, Illumination, and Implementation. Shaer et al. examined the potential of large language models to support collaborative ideation in two phases of creativity: the divergent phase as an idea generator, and the convergent phase as an idea evaluator~\cite{shaer2024ai}. Their findings suggest that GPT-4 can be used to provide unique and expanded perspectives, and evaluate ideas with similar qualities compared to human experts.

In this paper, we contributed a novel workflow to support human-AI co-creativity during prewriting. Inspired by prior work on human-human and human-AI co-creativity workflow, we enabled parallel collaboration with multiple heterogeneous microtasks processed by LLMs, which in our user study proved to afford more user agency, control, and creativity.

\subsection{Writing Support and Prewriting}
Writing support is a long-standing research topic in HCI. For decades a number of systems were designed to support different aspects of writing in varying phases, such as spelling and grammar checking \cite{leacock2010automated}, real-time summarisation \cite{dang2022beyond}, computer-generated feedback \cite{villalon2008glosser}, visualisation of topic flow \cite{o2009visualizing,o2011visualizing}, etc. For example, Dang et al. \cite{dang2022beyond} proposed a text editor that provides auto-updating summaries of paragraphs as margin annotations to help users plan, structure, and reflect on their writing. Iqbal et al. \cite{iqbal2018multitasking} proposed a micro-productivity tool that decomposed the complex task of document editing into smaller and simpler ``microtasks'' to enable users to edit Word documents via mobile devices. These work usually aimed to improve writing efficiency through various computer support.

There is a line of research that particularly focused on collaborative writing, in which the idea of ``microtask'' was constantly used. For example, to facilitate group collaboration, Teevan et al. \cite{teevan2016supporting} decomposed a writing task into three types of microtasks: idea generation, idea labeling and organisation, and writing paragraphs based on given ideas. Each microtask can be completed with limited awareness of the progress of others. Bernstein et al. \cite{bernstein2010soylent} proposed a tool that splits the task of writing and editing into stages and invites crowd workers to make suggestions, shorten, and proofread text.

Few writing tools cover the prewriting process, which often values both creativity and efficiency. Sadauskas et al. \cite{sadauskas2015mining} designed a dedicated system to support prewriting. Their system helps high school students prepare meaningful writing topics by mining their existing content on their social media. Lu et al. \cite{lu2018inkplanner} proposed a pen-and-touch-based tool to support prewriting via intelligent visual diagramming. Their system supports a set of diagram-based features such as the structuring of diagrams, automatic outline generation, etc.

Several creativity support tools (CSTs) were also developed to assist in generating writing ideas, although not explicitly framed as prewriting tools. For instance, Huang et al. \cite{huang2020heteroglossia} designed an add-on for Google Docs that enables writers to gather story ideas from the crowd using a role-play strategy. To support creative writing, Gero \cite{gero2019metaphoria} et al. designed a data-driven interface for writers to brainstorm metaphorical connections to an input word that are coherent with the context.

Our system contributes to HCI with an LLM-enhanced prewriting tool, which borrows the idea of ``microtask'' to support human-AI collaborative diagramming. The system aims to support pre-writing, and features a microtask management workflow to scaffold the collaborative diagramming process.
Many common features of prewriting and writing support (e.g., real-time summary \cite{dang2022beyond}, idea generation \cite{bernstein2010soylent}, argument elaboration \cite{uto2015academic}, etc.) can be achieved using our predefined microtasks, and users can also define their own microtasks to extend the system's capability for various purposes.

\subsection{Large Language Models to Support Writing}
Over the past few years, research in Artificial Intelligence (AI) has led to the development of increasingly powerful large language models (LLMs), including OpenAI's GPT-series \cite{radford2018improving} (Generative Pre-trained Transformer) such as ChatGPT and GPT-4 \cite{openai2023gpt4}. These LLMs demonstrate astonishing capabilities in handling a variety of complex Natural Language Processing (NLP) tasks.

With the development of LLMs, there has been an increasing interest in design goals beyond just efficiency, which has historically been the focus of writing assistance research \cite{jakesch2023co}. As such, there have been various design efforts to explore the potential of LLMs in supporting writing interface and interaction, including story writing \cite{singh2022hide}, text revision \cite{cui2020justcorrect, zhang2019type}, supporting inspiration \cite{inspiration1, lee2022coauthor, singh2022hide}, language abilities \cite{buschek2021impact}, or creative writing \cite{clark2018creative}. For example, Chung et al. proposed TaleBrush \cite{chung2022talebrush,chung2022talebrush_EA}, a system that enables users to co-create stories with LLMs via sketch interaction. Users can sketch how the protagonist's fortune should change chronologically in a story, and TaleBrush will generate story sentences and visualize the resulting fortune in the generation over the original sketch. Clark et al. \cite{clark2018creative} investigated the user experience of writing with language models by conducting case studies involving 46 participants given a slogan writing and a story writing task. Their findings reveal that users find unexpectedness in generations helpful and generally prefer a ``machine-in-the-loop'' mode where the language model plays a supporting role as a proactive writing partner.


It is worth noting that, randomness and unexpectedness in LLM generations can often be intrusive or cause breakdowns during human-AI collaborative writing \cite{clark2018creative}. Therefore, along this line of research, some literature was particularly dedicated to making LLM generations more controllable and explainable. For example, Wu et al. \cite{wu2022ai,wu2022promptchainer} proposed the idea of chaining LLM operations together. They break up complex writing tasks into smaller steps where the output of one or more steps can be used as input for the next.

Despite the potential benefits of using LLMs to support prewriting, there is still a paucity of research on effectively integrating LLMs into the prewriting process.
Recently, Jiang et al.~\cite{jiang2023graphologue} proposed a novel LLM augmented diagramming interface to decompose complex information in prewriting tasks.
Following their paradigm, we aim to transform the existing turn-taking workflow into a more efficient and creative alternative. \revision{It is similar to other diagrammatic interface such as ChainForge~\cite{arawjo2024chainforge,arawjo2023chainforge} or Sensecape~\cite{suh2023sensecape} that present multiple results simultaneously, but we allow AI agents for distinct purposes to proactively contribute visual diagrams in parallel.
To this end we use the notion of ``microtasking'' in human collaboration, which proves to be effective in our study.}
\section{Formative Study}
To understand how large language models could support diagramming-based prewriting, we conducted a formative study involving 10 participants with daily writing habits ranging from news articles to fiction writing. We used snowball sampling to recruit students in writing- or creativity-related majors (e.g., creative media, design, English literature etc.). All participants were ethnically Chinese, and English was their second language.

\subsection{Protocol}
After participants consented to the study, we asked them to develop a science fiction or thriller story plot using an LLM together with the traditional tools (a pen and a piece of paper) for diagramming or illustrating their ideas.
Participants accessed the LLM using the GPT-3 playground interface.
They were asked to think aloud during the prewriting process.
After completing the tasks, we asked participants to reflect on their experiences and strategies.

The whole process was audio-taped, screen-recorded, and later transcribed for analysis. Two coders performed thematic analysis \cite{corbin2014basics} of the transcripts with reference to the screen recording to extract collaboration strategies, patterns, and breakdowns. The coders later held a discussion to reach a consensus on the themes.

\subsection{Findings}
We report the findings of our formative study in two themes: human-LLM collaboration workflow and common challenges encountered. We found our participants already anthropomorphized the LLM as a collaborator, and delegated to it distinct tasks for both divergent and convergent thinking.

\subsubsection{Human-LLM Collaboration Workflow: Tasks and Initiative} \label{the_usage_of_LLMs_for_prewriting}
We observed that participants already implemented the parallel thinking strategy consciously or unconsciously.
They expected the LLM to perform multiple but distinctive functions, including generating additional ideas, elaborating on concepts, organizing fleeting thoughts, and enriching existing writing with details.
These functions were typically seen in creativity processes, covering both convergent and divergent thinking phases.

We found that users almost always preferred to take the initiative during the whole collaboration process, unless they ran out of ideas. They would use LLMs mostly to enrich their ideas with details, such as bridging a logical gap in a plot or providing a nuanced portrait of a scene.
Only when users had no initial ideas or hit writer's block, would they let the LLM take the initiative to generate ideas. P3 described the ideal role of LLM as similar to a ``\textit{second mind}'' that ``\textit{processes all the context in parallel}'' and ``\textit{provides ideas when requested}'', while he could still take control of the general prewriting process.

Meanwhile, we also noticed that users were generally tolerant and did not mind following the ideas in LLM output, which were often initially vague or confusing. Some users (P3-4, P8-9) were found to spend a long time iteratively refining their own prompts to improve the LLM's output.

\subsubsection{Challenges: Progress Tracking \& Communication Breakdown} \label{progress_tracking}
\paragraph{\underline{Progress Tracking}}
Many of our participants (P2-4, P6-7) particularly emphasized that tracking collaboration progress, and maintaining the ever-changing collaboration history while prewriting with the LLM was challenging. Because prewriting is iterative, and LLM generations can be random, our participants frequently needed to expand on specific points within a lengthy piece of writing in a new context, or re-generate content from a previous version. On such occasions, some participants (e.g., P2, P7, P10) mentioned they would need examples, suggestions, or templates as a reference to polish their prompts, and requested features to maintain the history of collaboration with LLM. 
\paragraph{\underline{Communication Breakdown}}
The uncertainty of prompt-based communication can often cause LLMs to generate unsatisfactory or even nonsensical results during prewriting. P2 and P4 reflected that, to effectively communicate with an LLM, proper and sufficient context should be articulated via prompts, which can be difficult in complex writing tasks. On these occasions, instead of rewriting their prompts, most participants (e.g., P1-4, P6, P9-10) chose to provide more context and ask LLMs to polish previous output. For example, P4 found the ending of an LLM-generated science fiction lacks originality. Instead of deleting the result and requesting a new one, she asked the LLM to avoid using banal superhero endings and explore existential questions, using an imperative sentence as if giving feedback to a human collaborator.

\subsection{Summary}
Our formative study reveals the creative nature of human-LLM collaboration during prewriting, where LLMs could offer additional perspectives and handle a range of complex ideation tasks. Although participants would like to take control most of the time, it is particularly beneficial that LLMs can run in parallel with users' diagramming activities and provide assistance when requested. During the collaboration, participants often found progress tracking and communication with the LLM using a conversational interface challenging. They requested features to support managing collaboration progress, and favoured an incremental feedback process to facilitate iteration.

\section{Design Goals}
Our primary design goal is to integrate LLMs into a diagramming-based interface to facilitate the application of the parallel thinking strategy in prewriting. We first introduce how we derive the core idea of ``microtasking'' to operationalize parallel thinking, and then report our three design goals to support a microtasking workflow.

\subsubsection*{\textbf{Microtasking for parallel thinking}}
The notion of ``parallel thinking'' separates the human thinking process into distinct functions and roles. 
In reality, parallel thinking can often be practiced through group collaboration, where each of the distinct roles can be played by different individuals or groups in parallel. We seek to simulate group collaboration to break down prewriting workflows into smaller, manageable, and independent tasks that support both divergent and convergent thinking processes.

To this end, we adopt the concept of ``microtasks'', commonly used in crowd sourcing \cite{latoza2014microtask,chen2017retool}, collaborative writing \cite{iqbal2018multitasking,birnholtz2013write,teevan2016supporting}, and group brainstorming \cite{chilton2019visiblends,teevan2016supporting}. Previous studies suggest that a microtasking workflow simplifies complex tasks \cite{cheng2015break,kokkalis2013taskgenies}, facilitates recovery from interruptions, and leads to higher quality of results \cite{cheng2015break}. Similarly, in a human-LLM collaboration scenario, we expect small manageable microtasks that require little context of one another to run concurrently can make complex prewriting tasks easier to coordinate, and save the need to iteratively articulate complex context in prompts.
Specifically, we derive the following three design goals to implement this idea. 

\newtcbox{\BboxL}{on line,
  colframe=brainstorm,colback=brainstorm,
  boxrule=0.5pt,arc=1.5pt,boxsep=0pt,left=3pt,right=3pt,top=3pt,bottom=3pt}
\newtcbox{\SboxL}{on line,
  colframe=summarise,colback=summarise,
  boxrule=0.5pt,arc=1.5pt,boxsep=0pt,left=3pt,right=3pt,top=3pt,bottom=3pt}
\newtcbox{\EboxL}{on line,
  colframe=elaborate,colback=elaborate,
  boxrule=0.5pt,arc=1.5pt,boxsep=0pt,left=3pt,right=3pt,top=3pt,bottom=3pt}
\newtcbox{\DboxL}{on line,
  colframe=draft,colback=draft,
  boxrule=0.5pt,arc=1.5pt,boxsep=0pt,left=3pt,right=3pt,top=3pt,bottom=3pt}
\newtcbox{\FboxL}{on line,
  colframe=freewrite,colback=freewrite,
  boxrule=0.5pt,arc=1.5pt,boxsep=0pt,left=3pt,right=3pt,top=3pt,bottom=3pt}
\newtcbox{\AboxL}{on line,
  colframe=associate,colback=associate,
  boxrule=0.5pt,arc=1.5pt,boxsep=0pt,left=3pt,right=3pt,top=3pt,bottom=3pt}

\renewcommand{\arraystretch}{0.75}

\renewcommand{\arraystretch}{0.75}
\begin{table*}[htb]
    \centering
    \resizebox{\linewidth}{!}{\begin{tabular}{lccl}
        \toprule
        \textbf{Microtask} & \textbf{Input Type} & \textbf{Output Type} &
        \textbf{Prompt} \\
        \midrule
        \renewcommand{\arraystretch}{2.5}
        \BboxL{\textcolor{white}{\textbf{Brainstorm}}}~\cite{huang2020heteroglossia,gero2019stylistic,wang2010idea,teevan2016supporting,lu2018inkplanner,wang2022interpretable} & keyword & keyword & 
        \makecell[l]{\underline{Find Related:} Brainstorm keywords related to [placeholder]. \\ \underline{Find Synonym:} Find synonyms for [placeholder].} \\
        & & & \\
        \SboxL{\textcolor{white}{\textbf{{Summarise}}}}~\cite{sadauskas2015mining,dang2022beyond} & sticky note & sticky note & \makecell[l]{\underline{TLDR:} Provide a TLDR version of the following:\textbackslash n[placeholder] \\ \underline{Top 3 keywords:} Summarise top 3 keywords of the following:\textbackslash n[placeholder]} \\
        & & & \\
        \EboxL{\textcolor{white}{\textbf{Elaborate}}}~\cite{sadauskas2015mining,uto2015academic,jeon2021fashionq} & concept & concept & \makecell[l]{\underline{Provide Examples:} What are examples of [placeholder]. \\ \underline{Clarification:} Provide a simple explanation of [placeholder].} \\
        & & & \\
        \DboxL{\textcolor{white}{\textbf{Draft}}}~\cite{lu2018inkplanner,chung2022talebrush} & section & sticky note & \makecell[l]{\underline{Abstract:} [placeholder]\textbackslash n\textbackslash nWrite an abstract of the above outline. \\ \underline{Overview:} [placeholder].\textbackslash n\textbackslash nWrite an overview of the above outline.} \\
        & & & \\
        \FboxL{\textcolor{white}{\textbf{Freewrite}}}~\cite{baroudy2008procedural,lu2018inkplanner} & sticky note & sticky note & \makecell[l]{\underline{Co-creation:} [placeholder].\textbackslash n Continue to write.} \\
        & & & \\
        \AboxL{\textcolor{white}{\textbf{Associate}}}~\cite{gero2019metaphoria,chilton2019visiblends,wang2021popblends,faste2012untapped,hope2022scaling,wang2022interpretable} & nodes & keyword &  \makecell[l]{\underline{Find Relationship:} Clarify the relationship between [placeholder] and [placeholder] in simple words.} \\
        \bottomrule
    \end{tabular}}
    \caption{Six default microtasks of \textit{Polymind}}
    \label{table:microtasks} 
\end{table*}

\subsubsection*{\textbf{Goal 1: Scaffold visual-diagramming-based prewriting with microtasks}}
We aim to scaffold the prewriting process with a diagramming tool that supports common strategies such as concept mapping, mind mapping, outlining, etc. To enable natural collaboration with an LLM while diagramming, we are inspired by the concept of ``macros'' \cite{kurlander1992history} and seek to address the uncertainty of collaboration goals by defining default microtasks, and allowing users to customise their requirements or rapidly delegate their own microtasks.

Our formative study inspired us to draw upon existing literature on creativity support tools (CSTs) beyond prewriting itself to define default microtasks, as the collaboration workflow appeared to be a typical creativity process. We conducted a survey of both CST and writing tool literature by searching two academic databases, ACM Digital Library and Google Scholar, using three keywords: ``writing'', ``prewriting'', and ``creativity support''. We reviewed the top 100 entries for each keyword.
Based on this survey, we identified 6 microtasks: ``Brainstorm'', ``Elaborate'', ``Summarise'', ``Draft'', ``Freewrite'', and ``Associate'', as summarised in \autoref{table:microtasks}. Of these microtasks, ``Brainstorm'' and ``Associate'' are typical divergent thinking tasks seen in existing CSTs (e.g., conceptual blending of \cite{wang2021popblends,chilton2019visiblends}, attribute detection of \cite{jeon2021fashionq}, group ideation of \cite{teevan2016supporting,wang2010idea}, etc.)
``Draft'' and ``Freewrite'' are common features in prewriting tools \cite{lu2018inkplanner,sadauskas2015mining}. ``Elaborate'' and ``Summarise'' are commonly used in the literature of writing support~\cite{uto2015academic,dang2022beyond}, as convergent thinking tasks to help articulate or organise existing ideas.

\subsubsection*{\textbf{Goal 2: Facilitate task management}}
Task management is crucial in human collaboration. In our scenario, a user acts as a leader who determines the goals and progress of the collaboration. Therefore, she should be granted sufficient control to manage microtasks. To this end, we further derive three sub-goals from the literature on human collaboration and our formative study.
\paragraph{\textbf{Goal 2.1: Provide awareness}}
The awareness information about other collaborators while using groupware \cite{gutwin2002descriptive} is vital in tasks such as collaborative writing \cite{birnholtz2013write} and collaborative learning \cite{fransen2011mediating}. On a prewriting interface (e.g., a diagramming canvas), where elements can be loosely organised and often scattered around, it might be hard to notice other collaborators' operations without proper design support. Therefore we seek to provide awareness features so that users can easily track the status of each microtask, and the results returned by each microtask.
As informed by ~\cite{gluck2007matching}, we aim to design different levels of awareness features that match the utility of different interruption types.

\paragraph{\textbf{Goal 2.2: Support progress tracking}}
Progress tracking is an essential aspect of many collaboration tasks, especially collaborative writing \cite{birnholtz2013write,birnholtz2012tracking}. Our formative study suggests that it is also a concern while collaborating with an LLM. We aim to help users manage the results of each microtask in a less demanding way, so that they do not clutter users' diagrams but can be merged into them once accepted.
\paragraph{\textbf{Goal 2.3: Facilitate human feedback}}
Feedback is essential for improvements \cite{dow2011shepherding,haug2021feeasy,huang2018feedback}. In our formative study, users generally preferred feedback-like communication upon unsatisfactory generations. Therefore we aim to facilitate user feedback to enhance LLM-generated content. In our system, the ``feedback'' is provided to an LLM, which implies that it should convert user requirements into actionable prompts.

\subsubsection*{\textbf{Goal 3: Apply mixed initiative}}
Although users preferred to maintain control most of the time in our formative study, they still wanted the LLM to take the initiative when running out of ideas, a common hurdle during prewriting. Sometimes, they even expected the LLM to further clarify or improve its output. Therefore, we aim to apply the principle of ``mixed initiative'' \cite{horvitz1999principles}, and allow a microtasking LLM to infer the focus of attention of the user to determine the timing of suggestions. To keep users in control and minimize the cost of inference errors, we aim to enable users to manage the initiative of each individual microtask. 
\newtcbox{\BboxS}{on line,
  colframe=brainstorm,colback=brainstorm,
  boxrule=0.5pt,arc=1pt,boxsep=0pt,left=2pt,right=2pt,top=2pt,bottom=2pt}
\newtcbox{\SboxS}{on line,
  colframe=summarise,colback=summarise,
  boxrule=0.5pt,arc=1pt,boxsep=0pt,left=2pt,right=2pt,top=2pt,bottom=2pt}
\newtcbox{\EboxS}{on line,
  colframe=elaborate,colback=elaborate,
  boxrule=0.5pt,arc=1pt,boxsep=0pt,left=2pt,right=2pt,top=2pt,bottom=2pt}
\newtcbox{\DboxS}{on line,
  colframe=draft,colback=draft,
  boxrule=0.5pt,arc=1pt,boxsep=0pt,left=2pt,right=2pt,top=2pt,bottom=2pt}
\newtcbox{\FboxS}{on line,
  colframe=freewrite,colback=freewrite,
  boxrule=0.5pt,arc=1pt,boxsep=0pt,left=2pt,right=2pt,top=2pt,bottom=2pt}
\newtcbox{\AboxS}{on line,
  colframe=associate,colback=associate,
  boxrule=0.5pt,arc=1pt,boxsep=0pt,left=2pt,right=2pt,top=2pt,bottom=2pt}

\newtcbox{\CboxS}{on line,
colframe=custom,colback=custom,
boxrule=0.5pt,arc=1pt,boxsep=0pt,left=2pt,right=2pt,top=2pt,bottom=2pt}

\newtcbox{\CCboxS}{on line,
colframe=custom2,colback=custom2,
boxrule=0.5pt,arc=1pt,boxsep=0pt,left=2pt,right=2pt,top=2pt,bottom=2pt}
  
\section{Use Case Scenario}
In this section, we illustrate the workflow of \textit{Polymind} via a use case scenario. Suppose that Bob, an HCI researcher, would like to use fictional narratives to promote his research on social media. He therefore chooses to use \textit{Polymind} to plan his fiction writing.

\subsection*{Task Management}
Bob starts with the key insights of his paper, that parallel inputs from AI agents such as LLMs can significantly increase writers' creativity. His primary goals of using \textit{Polymind} are brainstorming narrative lines, and working out a rough outline. Therefore he keeps three microtasks in the proactive mode to quickly expand his ideas: \BboxS{\textcolor{white}{Brainstorm}}, \EboxS{\textcolor{white}{Elaborate}}, and \AboxS{\textcolor{white}{Associate}}; and switched all other microtasks to the reactive mode so that they will not be intrusive.

\subsection*{Collaborative Brainstorming}
To get some initial ideas, Bob uses \textit{Polymind} to perform mind mapping, which focuses on tracking spontaneous and free-form ideas, and their associations.
He first creates some keywords and concepts on the canvas: such as ``parallel collaboration'', ``creative writing'', etc. The three proactive microtasks, \BboxS{\textcolor{white}{Brainstorm}}, \EboxS{\textcolor{white}{Elaborate}}, \AboxS{\textcolor{white}{Associate}} shortly returns some related keywords, example scenarios, and associations between these diagrams.

Bob accepts three diagrams: ``synchronous tasks'', ``mutual goals'', and ``flash fiction''. These remind him of a story where a former fiction writer revolutionizes the fiction writing industry by leveraging multiple robots in an assembly line to mass produce flash fictions. Each robot is configured to handle a distinct microtask on the assembly line, working synchronously towards a common goal.

\subsection*{Task Delegation}
At this point, Bob feels he has obtained some concrete ideas, and would like to organize them via concept mapping.
This method aims to outline structures and relationships between concepts. He also wants to ask the LLM how to make a story more engaging. Therefore, he delegates a new microtask \CboxS{\textcolor{white}{Improve}}, which returns suggestions for improvements.
He leaves it in the \textit{reactive} mode so that it can provide suggestions based on the ever-changing diagram upon request.

\subsection*{Idea Clarification}
Bob starts concept mapping by specifying elements of the story. He creates some diagram nodes, such as ``Character: Fiction Writer \& Robots'', ``Event: Revolutionizing the industry through assembly lines of flash fictions'', etc. He then creates a section over these diagrams to group them together, and uses the microtask \DboxS{\textcolor{white}{Draft}} to request an outline, and clicks the \FboxS{\textcolor{white}{Freewrite}} microtask several times to continue writing to see different endings.

Bob feels that these results are somewhat bland, and therefore asks the microtask \CboxS{\textcolor{white}{Improve}} to suggest improvements based on these diagrams. The results suggest adding to the beginning the conflict between the fiction writer and his former boss that fired him for lacking creativity. It reminds Bob that he could depict the former boss as a firm advocate of turn-taking conversational robots, and unveil the superiority of a parallel collaboration through the main character's adventure.

\begin{figure}[ht]
\includegraphics[width=\linewidth]{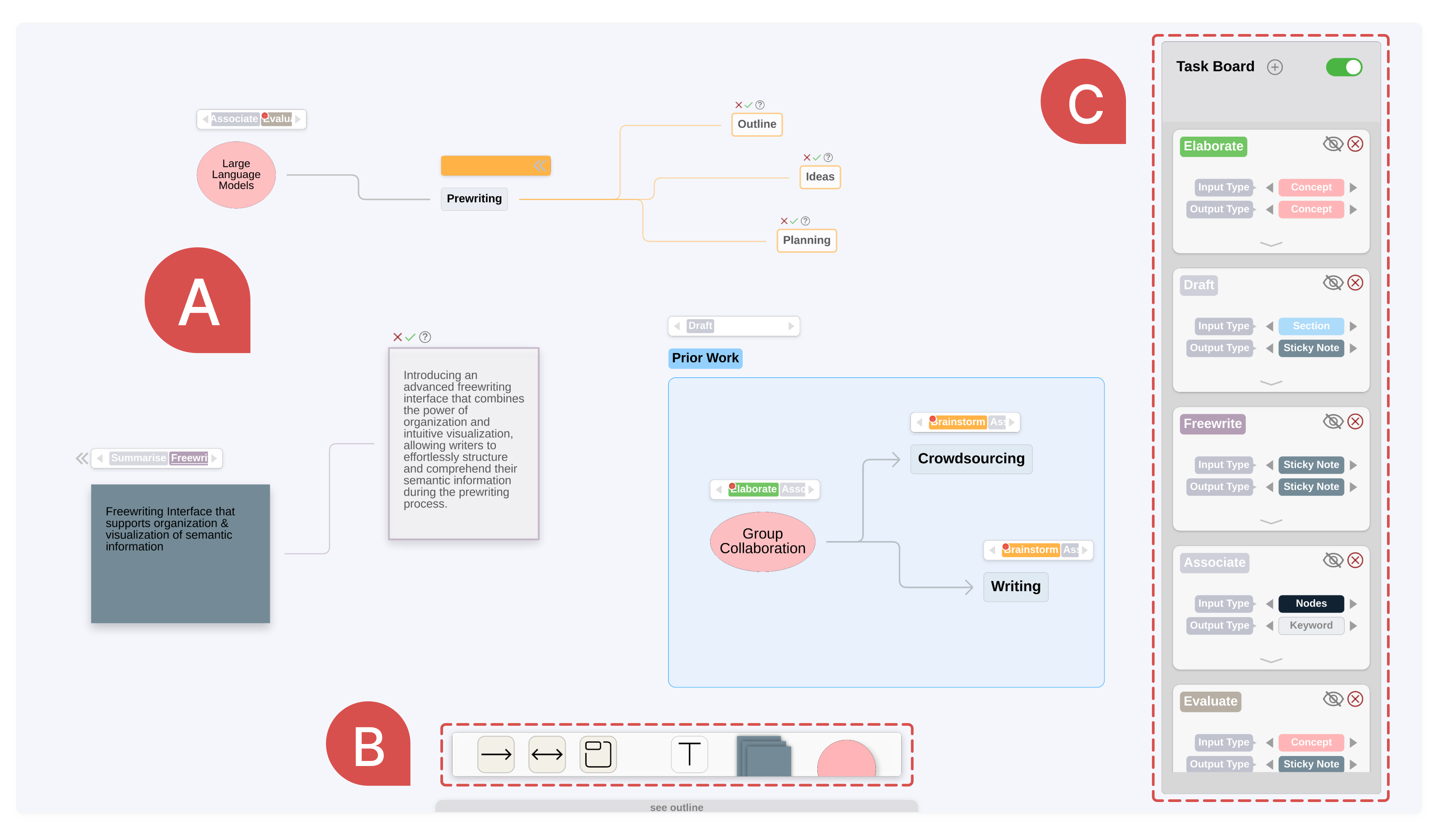}
\caption{The interface of \textit{Polymind} comprises: A. a diagramming canvas B. a toolbar C. a task board}
\label{fig:UI}
\end{figure}
\section{Designing Polymind}
\revision{
\textit{Polymind}'s interface provides a range of diagramming features commonly used in prewriting strategies, and a ``task board'' overlaid on the canvas for microtask management, as shown in ~\autoref{fig:teaser} and ~\autoref{fig:UI}. We map the input and output of each microtask to diagram types on the canvas. To facilitate collaboration and task management, we introduce the notion of ``task header'', and ``task cards''. The former displays notifications and previews of microtask results on a specific diagram, while the latter displays specifications of a microtask on the ``task board''.
To apply mixed initiative, we define two initiative modes: proactive and reactive. We also design a set of workflows to provide awareness information.
In this section, we first provide an overview of the \textit{Polymind} interface, and then elaborate on each of our key features.
}


\begin{figure}[ht]
\centering
\includegraphics[width=\linewidth]{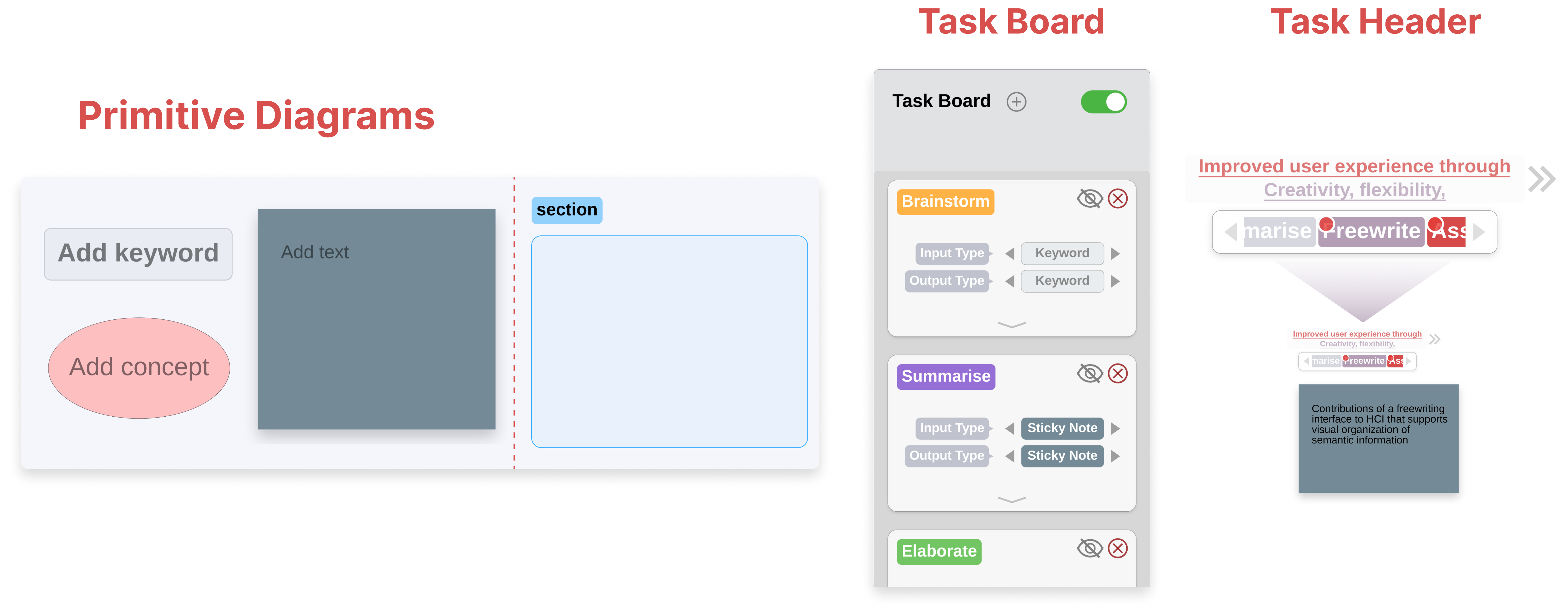}
\caption{\textit{Polymind} supports three basic diagrams (or nodes), and allows users to draw a section over diagrams. The task board interface supports microtask management, and maps microtask input and output to different diagram types on the canvas. The results of microtasks are displayed as notifications and previews on the task header before being expanded and accepted.}
\label{fig:diagrams}
\end{figure}

\subsection{Main Interface}
The interface of \textit{Polymind} comprises a diagramming canvas, a task board, and a toolbar (\autoref{fig:UI}).
\revision{The diagramming canvas includes all diagrams and their topological connections. We defined three basic diagrams: \textbf{\textcolor{keyword}{\textit{keyword}}}, \textbf{\textcolor{concept}{\textit{concept}}}, \textbf{\textcolor{sticky_note}{\textit{sticky note}}}, each with a unique shape affording varying text lengths (keywords the shortest, and sticky notes for the longest pieces of text). Users can connect diagrams through directed or undirected edges, or draw a \textbf{\textcolor{section}{\textit{section}}} over diagrams (see ~\autoref{fig:diagrams}).

The task board displays all microtasks and their specifications in distinct ``task cards'', each including input and output types, prompts, etc. The results of each microtask are displayed as notifications and previews on the ``task header'' over that processed diagram before they are expanded and accepted (see ~\autoref{fig:task_status}).}

\subsubsection{The Diagramming Canvas}
The diagramming canvas supports most common diagrams. Over each of these diagrams, a task header is attached to display the status of microtasks. All resulting diagrams of microtasks are displayed using a hollow shape in contrast to user created diagrams.

\paragraph{\underline{Diagrams}}
To scaffold the diagramming process for \textit{Goal 1}, we define three primitive diagrams that are commonly used in diagramming or prewriting tools (e.g., Inkplanner \cite{lu2018inkplanner}, Figma \footnote{https://www.figma.com/}, etc.): \textbf{\textcolor{keyword}{\textit{keyword}}} (displayed as text labels), \textbf{\textcolor{concept}{\textit{concept}}} (represented by an ellipse), and \textbf{\textcolor{sticky_note}{\textit{sticky note}}} (see \autoref{fig:diagrams}). We use sizes, shapes and the placeholder text upon creation (``Add Keyword'', ``Add Concept'', ``Add text'') to guide users to type text of different lengths and of different functions into different diagrams (e.g, brief words in \textbf{\textcolor{keyword}{\textit{keyword}}}, short phrases in \textbf{\textcolor{concept}{\textit{concept}}}, long paragraphs in \textbf{\textcolor{sticky_note}{\textit{sticky note}}}), so that we can properly define microtasks handling various types of input to support all three stages of LLM usage (see \ref{the_usage_of_LLMs_for_prewriting}).

These three primitive diagrams can be selected, moved, resized or scaled as per users' needs. Additionally, users can establish two types of connection between these diagrams: directed (arrow) or undirected (line). We also allow users to create a \textbf{\textcolor{section}{\textit{section}}} (\autoref{fig:diagrams}) among these diagrams, and assign titles to these sections. On such an interface, users can perform most diagram-based prewriting strategies such as mind mapping, concept mapping, argument mapping, etc.

It is important to note that the diagramming canvas maintains a graph-like (usually tree-like) structure, and we later refer to those primitive diagrams as ``nodes'' at times for simplicity.

\paragraph{\underline{The Task Header}}
To support \textit{Goal 2} (especially \textit{Goal 2.1}) and \textit{Goal 3}, we design a task header that is attached to each diagram (i.e., \textbf{\textcolor{keyword}{\textit{keyword}}}, \textbf{\textcolor{concept}{\textit{concept}}}, \textbf{\textcolor{sticky_note}{\textit{sticky note}}}, and \textbf{\textcolor{section}{\textit{section}}}) on the diagramming canvas to display the status of each microtask on the diagram element (\autoref{fig:task_status}). On the task header, the name of each microtask is displayed as small text labels. By default, all microtasks are activated (see \ref{task_initiative}) and filled with distinctive colours. Each task header also has a preview panel (\autoref{fig:notifications_and_previews}) that displays key points of unread microtask results. It will only pop up when users hover over the task header, and there are unread microtask results.

\begin{figure}[h]
\begin{minipage}{.5\textwidth}
    \captionsetup{width=\linewidth}
    \vspace{0pt}
    \includegraphics[width=\linewidth]{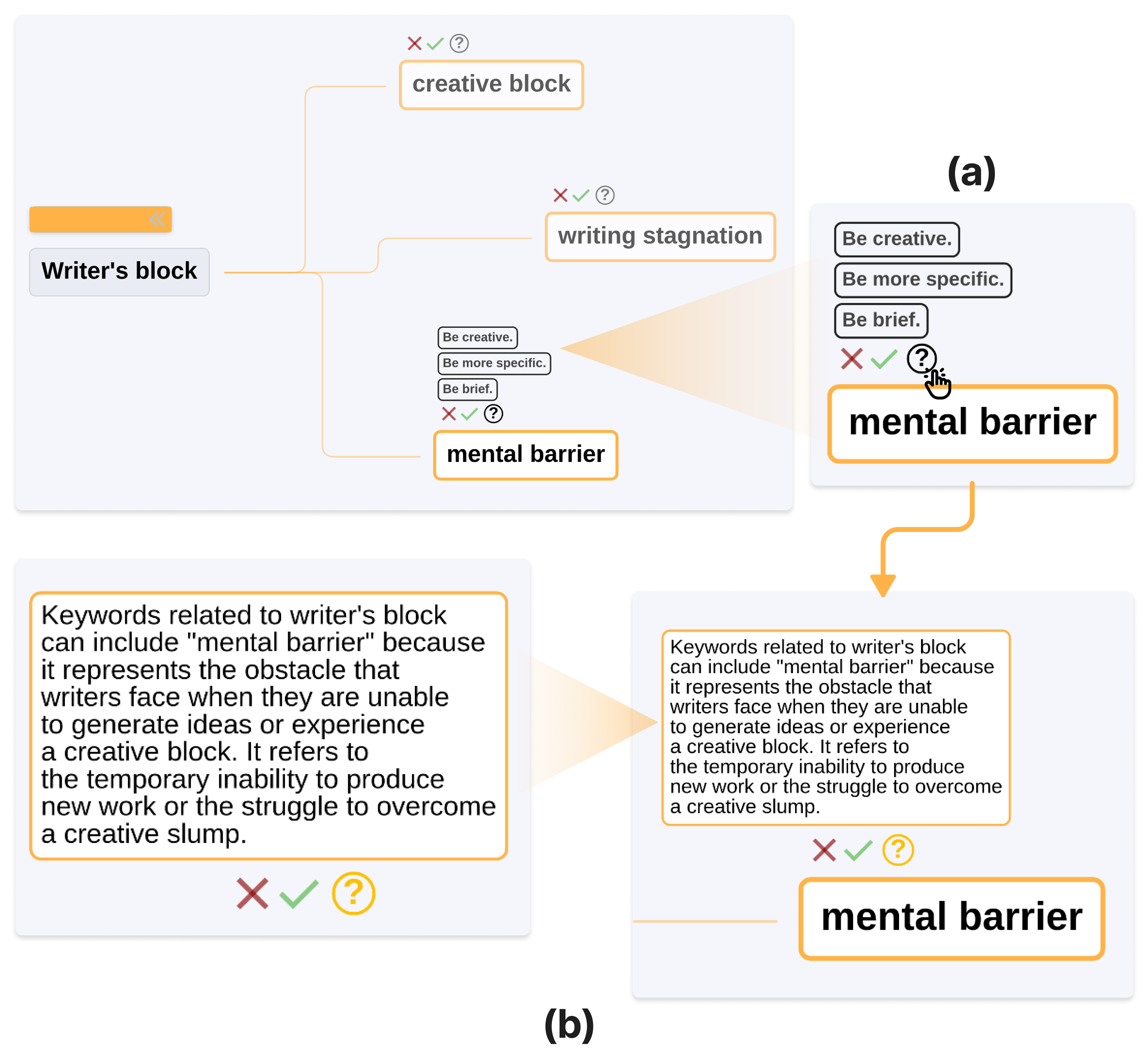}
    \caption{(a) When the user hovers over the three icons over the resulting diagram, suggestions to improve the generation will pop up. The user can click on these suggestions to request a regeneration. (b) If a user clicks on the question mark icon, the system will return an explanation of the generated result.}
    \label{fig:feedback}
\end{minipage}
\begin{minipage}{.45\textwidth}
    \captionsetup{width=.9\linewidth}
    \vspace{0pt}
    \includegraphics[width=\linewidth]{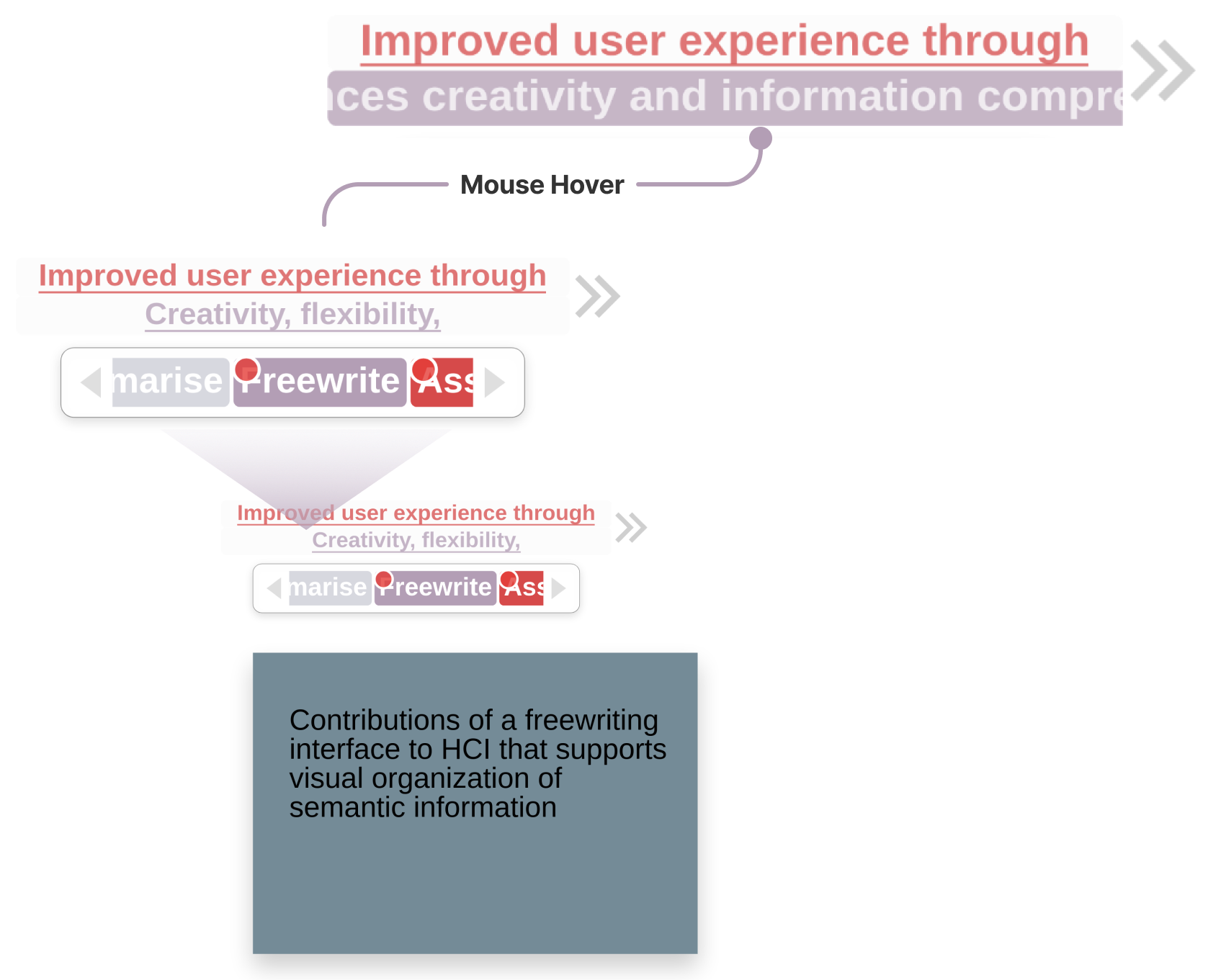}
    \caption{Once the mouse hovers over key points of a microtask on the preview panel for 1.5 seconds, the system will present a summary of the generated results using a news ticker effect.}
    \label{fig:preview_summary}
\end{minipage}
\end{figure}

\paragraph{\underline{Results of Microtasks}} To support \textit{Goal 2.2} we display all resulting diagram nodes using a hollow shape (filled in white), including \textbf{\textcolor{keyword}{\textit{keyword}}}, \textbf{\textcolor{concept}{\textit{concept}}}, and \textbf{\textcolor{sticky_note}{\textit{sticky note}}}. The border colour indicates the specific microtask each resulting node belongs to. A user can choose to discard or accept a resulting node; upon acceptance it will be displayed normally, the same as user-created diagram nodes.

Apart from the two icons for accepting or discarding the diagram node, to support \textit{Goal 2.3}, we design a ``question mark'' icon (see \autoref{fig:feedback}) that users can click on to request an explanation for this specific generation. For each resulting node, we also provide three heuristic suggestions for users to request a regenerated node, ``Be creative'', ``Be more specific'', and ``Be brief''. These suggestions will pop up if users hover over those three icons (see \autoref{fig:feedback}).

\subsubsection{The Task Board}
The task board of \textit{Polymind} enables users to configure and delegate microtasks, to support our \textit{Goal 2}. The task board borrows the design of Trello \footnote{https://trello.com/} board, which consists of a list of ``task cards''. Users can click on the ``add'' icon to delegate a new microtask, or toggle the ``visibility'' switch on the upper-right corner to hide all task headers on the diagramming canvas.

\begin{figure}[t!]
\centering
\begin{subfigure}{.32\linewidth}
    \centering
    \includegraphics[width=\linewidth]{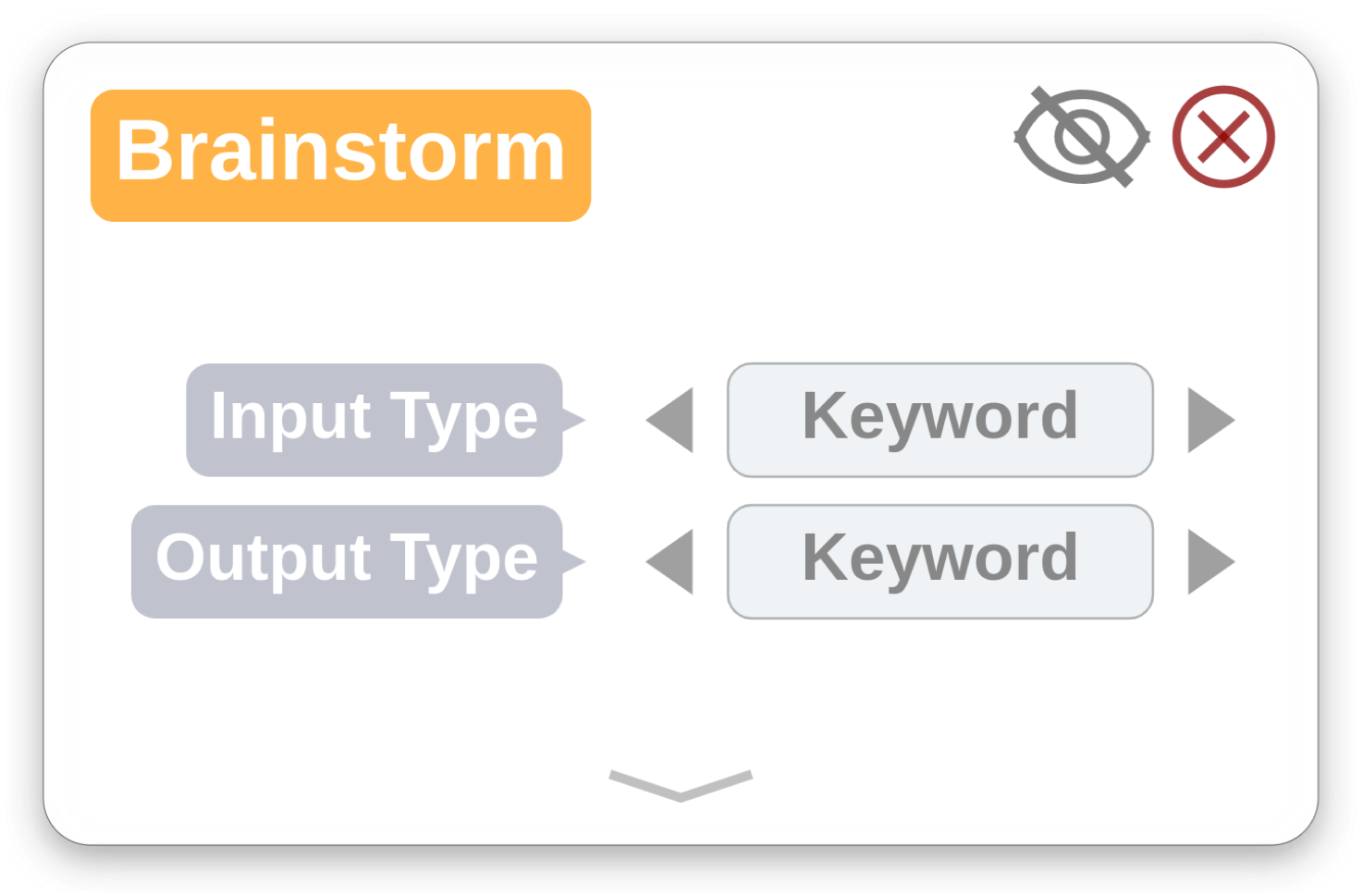}
    \subcaption{The first page}
    \label{fig:task_card_page1}
\end{subfigure}
\begin{subfigure}{.32\linewidth}
    \centering
    \includegraphics[width=\linewidth]{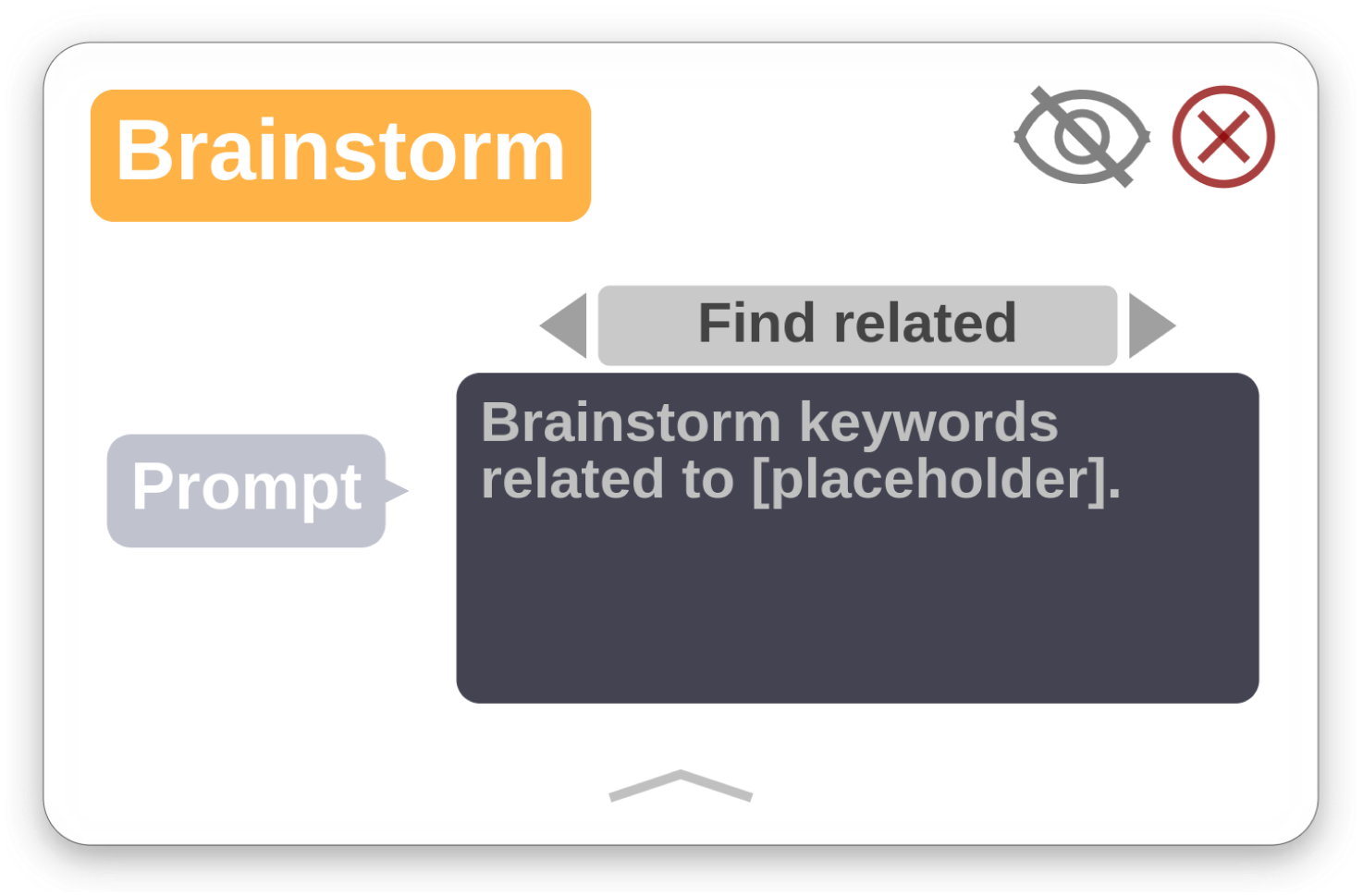}
    \subcaption{The second page}
    \label{fig:task_card_page2}
\end{subfigure}
\begin{subfigure}{.32\linewidth}
    \centering
    \includegraphics[width=\linewidth]{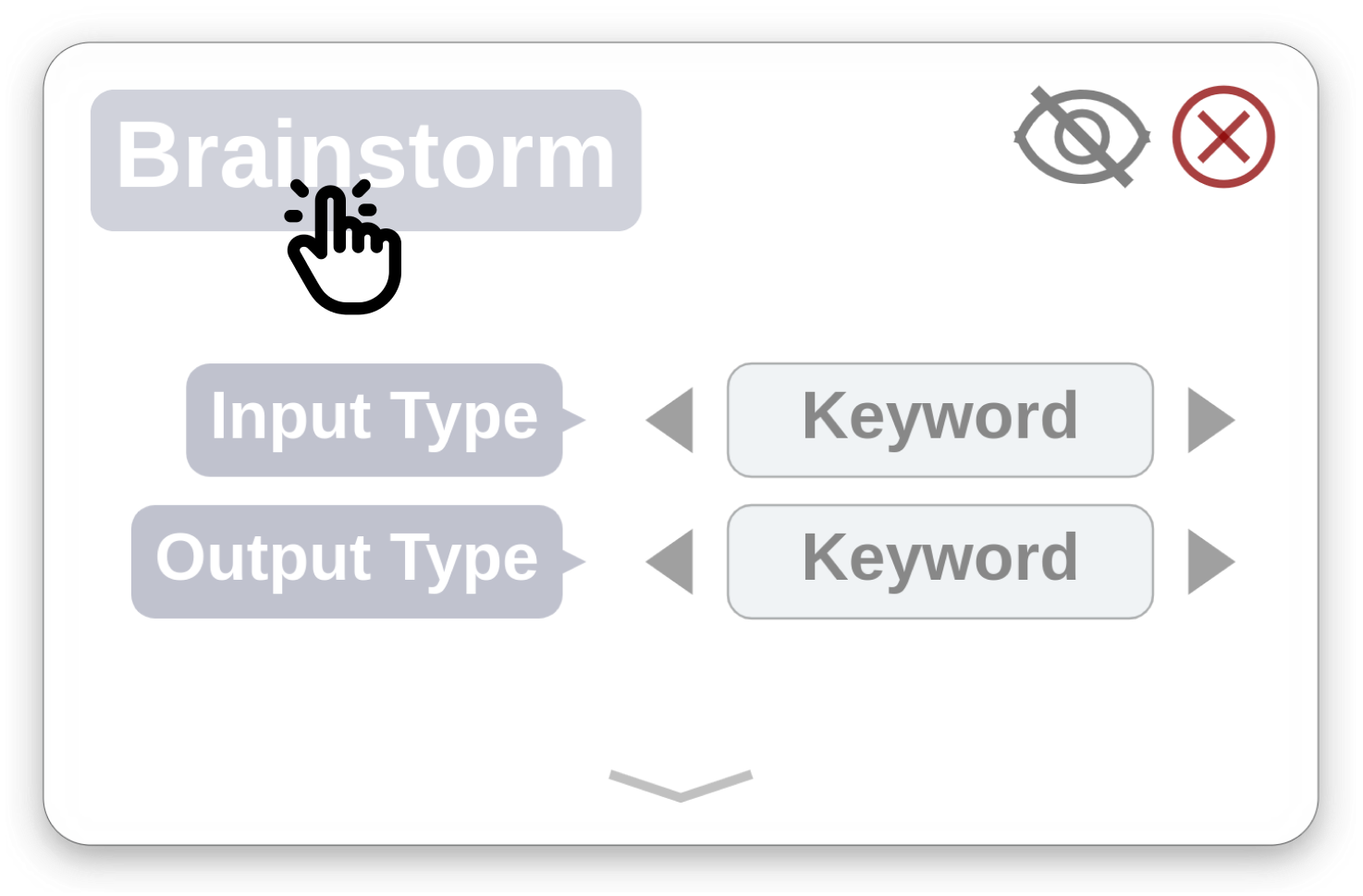}
    \subcaption{Toggling initiative modes}
    \label{fig:task_card_inactive}
\end{subfigure}
\caption{The task card of a microtask. The text label of the task name is indicated with a distinctive colour. (a) By default, the card shows information of 
its input and output types, which are configurable. (b) Users can click on the arrow at the bottom of the card to turn pages, and edit prompts in the second page. The user can also switch between predefined prompts. (c) The user can click on the text label to toggle initiative modes globally, or the ``visibility'' icon to display or hide all resulting diagrams.}
\label{fig:task_card}
\end{figure}

\paragraph{\underline{Task Card}}
A task card contains specifications of a microtask, including the task name, input type, output type, initiative mode (global), visibility (global), and prompts to communicate with the LLM (\autoref{fig:task_card}). The name of the microtask is always displayed in the upper-left corner, each indicated by a distinctive colour (same as resulting diagrams or labels on task headers).

Users can browse through task specifications by clicking on the ``page turn'' icon at the bottom; delete the microtask by clicking on the ``delete'' icon; toggle visibility of all resulting diagrams on the canvas by clicking on the ``visibility'' icon; toggle the initiative modes (see \ref{task_initiative}) by clicking on the task name (the text label). They can also change the specifications of each microtask (see \ref{task_configuration}).

\paragraph{\underline{Default Microtasks}} There are 6 predefined microtasks in \textit{Polymind}, which are \BboxS{\textcolor{white}{Brainstorm}}, \EboxS{\textcolor{white}{Elaborate}}, \SboxS{\textcolor{white}{Summarise}}, \DboxS{\textcolor{white}{Draft}}, \FboxS{\textcolor{white}{Freewrite}}, and \AboxS{\textcolor{white}{Associate}}, as specified in our \textit{Goal 1}. Each microtask has a default input and output type, and prompt templates to communicate with the LLM. In practice, when operating on an input node, each microtask will replace the ``[placeholder]'' with the text in that node to prompt the LLM. For details, we refer our readers to \autoref{table:microtasks}. These defaults can all be changed later as per users' needs.

For the input type of ``nodes'', the microtask will, given a node on the canvas, sample another nearby node to perform an operation, and generated diagrams will be linked to both nodes. For the ``section'' input, the microtask will calculate the outline of all nodes (similar to \cite{lu2018inkplanner}) within the section by performing a depth-first search (DFS) of all non-leaf nodes. For example, for a section with a standalone keyword ``creativity'', and a root node ``writing'' connecting to another two keywords, ``drafting'', ``proofreading'', the resulting DFS sequence will be:
\begin{quote}
Writing

 -\ \ \ Drafting
 
 -\ \ \ Proofreading
 
 Creativity
\end{quote}

\subsubsection{The Toolbar}
The toolbar comprises four icons, a pile of sticky notes, and an ellipse representing a concept (see \autoref{fig:teaser}). Users can drag a sticky note or a concept (the ellipse) from the toolbar to the diagramming canvas, or click on the ``text'' icon and then click on the canvas again to add a keyword.
The two leftmost icons are used for connecting diagram nodes using arrows (directed) and lines (undirected). Users can click on the icon and select anchor points of a node to connect with another. There is also an icon for sectioning where users can click on it, and draw a rectangle over diagrams as a section.

\subsection{Processing a Proactive Microtask}
To achieve our \textit{Goal 3}, we introduce two initiative modes: proactive and reactive (\autoref{fig:task_status}). In the reactive mode, the microtask will function similarly to a button, and will not execute until a user clicks on its label on a task header. Those reactive microtasks will be displayed using a gray colour on the task header. By default, all microtasks are activated and run automatically in parallel with user activities.
\revision{
\minor{We sample every \textit{x} (in practice, $x=5$) seconds a diagram node for each input type: \textbf{\textcolor{keyword}{\textit{keyword}}}, \textbf{\textcolor{concept}{\textit{concept}}}, \textbf{\textcolor{sticky_note}{\textit{sticky note}}}, and \textbf{\textcolor{section}{\textit{section}}}.}
Each microtask will operate on the sampled diagram corresponding to its input type, if it does not have unchecked notifications and is not displaying results on that diagram.
}
Results of these microtasks will be displayed as notifications and previews before users manually expand all resulting diagrams.

\subsubsection{Inferring User Attention}
In an earlier pilot study of our system, we found that users might lose track of generation results if a microtask was constantly operating on diagram nodes far away from users' focus of attention. Therefore, for our \textit{Goal 2.1}, we want each proactive microtask to infer the user's attention and mainly operate on diagrams of a user's focus. We make the assumption that diagram nodes that can be most easily selected are nodes of users' focus. In other words, the wider the node, or the closer to the mouse cursor, the more likely the node is the focus of users' attention, and operating on it will more likely make users aware. In practice, every \textit{x} seconds, for each node among each of the 5 types of input, we compute an index of difficulty ($ID$) according to Fitts' law \cite{fitts1954information}:
\begin{equation*}
    ID = \log_2 \frac{2D}{W}
\end{equation*}
where $D$ denotes the distance between the current mouse position and the node centre, and $W$ denotes the width of the node.

However, if we choose the node with the lowest $ID$ each time, we will end up always choosing the same node for operation if a user barely moves the mouse within a period of time. Therefore, in practice, we sample a node from a uniform distribution where the probability of a node $i$ being sampled as an input type $T$ is computed as:
\begin{equation*}
p(i, T) = \frac{ID_i}{\sum_{j \in T} ID_j}
\end{equation*}
$T$ includes \textbf{\textcolor{keyword}{\textit{keyword}}}, \textbf{\textcolor{concept}{\textit{concept}}}, \textbf{\textcolor{sticky_note}{\textit{sticky note}}}, and \textbf{\textcolor{section}{\textit{section}}}.
\revision{For microtasks that operate on two \textbf{\textit{nodes}} (e.g., the default \AboxS{\textcolor{white}{Associate}}), we first sample a primitive diagram, and then randomly sample a nearby diagram for prompting.}

\subsubsection{Notifications and Previews of Microtask Results}
We use two levels of attention draw features: notifications and previews of microtask results to support \textit{Goal 2.1}, so that results ready for display will not intrude users' diagramming process.

\begin{figure*}[ht]
\includegraphics[width=\textwidth]{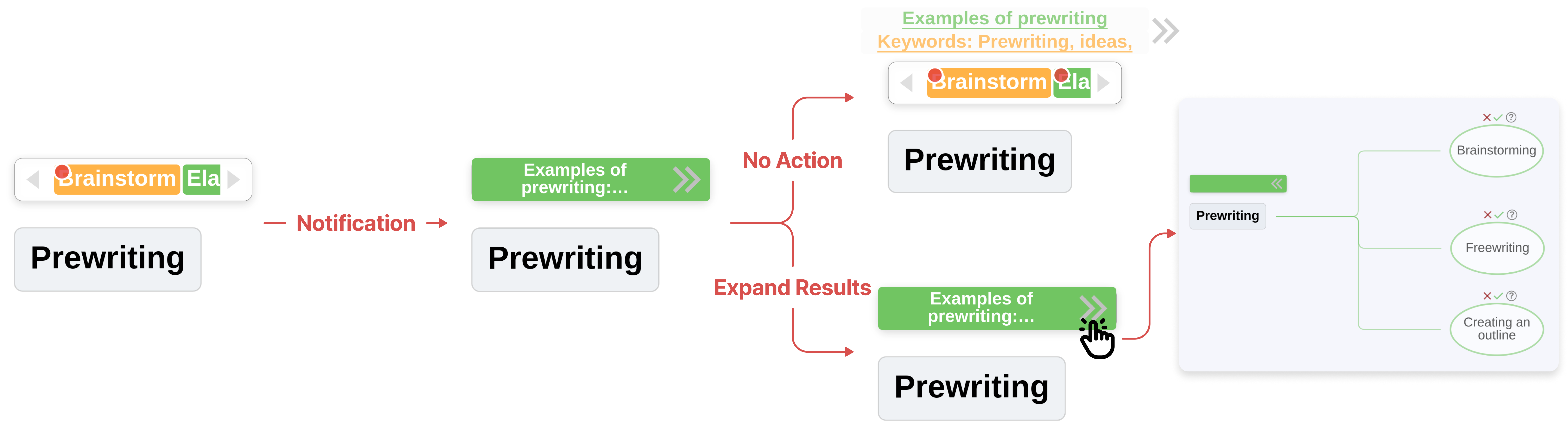}
\caption{The processing of a proactive microtask. Once results are obtained from the LLM, it will first ``draw a curtain'' over the task header to notify users, displaying key points of the generation. Users can click on the ``expand'' icon on the ``curtain'' for a quick view. If a user does not expand these result, it will be marked as unread notifications. Once users hovers over the task header, a preview panel will pop up, showing a preview of all unread results.}
\label{fig:notifications_and_previews}
\end{figure*}

\paragraph{\underline{Notifications}}
To notify the user once results of a proactive microtask are ready, it will first ``draw a curtain'' (see \autoref{fig:notifications_and_previews}) over the task header of the node being operated. The curtain will be filled with a distinctive colour indicating which microtask those results are from, and meanwhile, display key points summarising the generated results. Users can click on the ``expand'' icon on the right to display results, and then click on the ``collapse'' icon to hide them. If the user does not perform any operations, the curtain will collapse, and those results will be marked as ``unread'' with a small red circle on the upper-left corner as a sign of notification. (the red circle in \autoref{fig:notifications_and_previews}).

\paragraph{\underline{Previews}}
Once there are unread results, the preview panel of a task header will store a preview of these results. When a user hovers over the task header, the preview panel (should there be any results unread) will pop up, displaying key points of results of each microtask, using labels filled with distinctive colours (see \autoref{fig:notifications_and_previews}). Once a user hovers over a specific label for 1.5 seconds, the key point will turn into a brief summary of the microtask results, displayed using effects resembling a news ticker (see \autoref{fig:preview_summary})

\subsection{Managing Microtasks}
To support our \textit{Goal 2} and \textit{Goal 3}, we design a set of features to support configuring existing microtasks, and delegating new microtasks.

\begin{figure*}[ht]
\centering
\includegraphics[width=\textwidth]{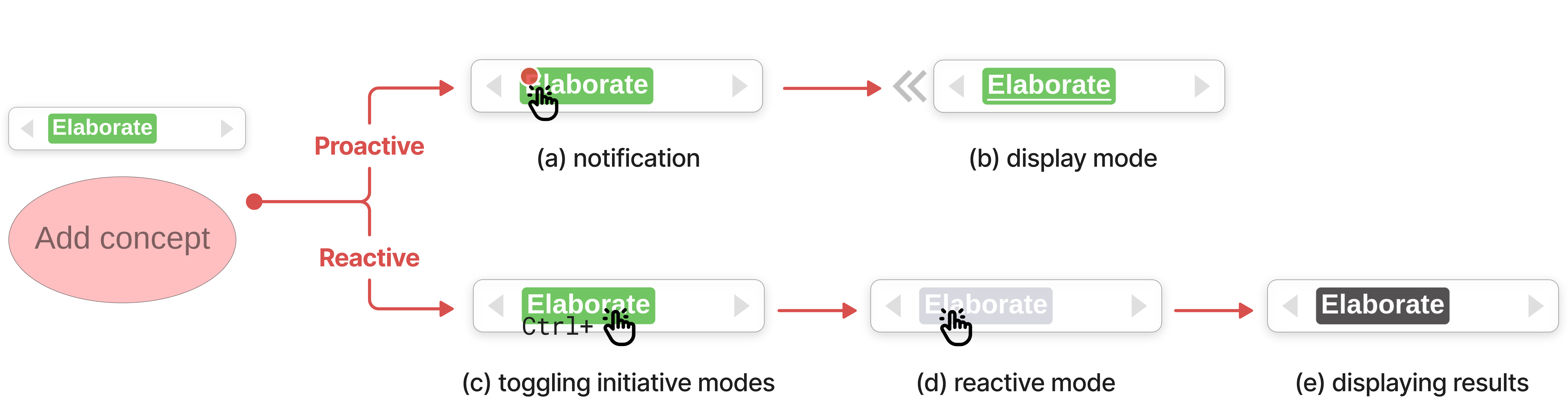}
\caption{Different status of a microtask on a particular node.}
\label{fig:task_status}
\end{figure*}
\subsubsection{Toggling Visibility}
For the purpose of \textit{Goal 2.2}, users can toggle the visibility of microtask results both globally and locally. By clicking on the ``visibility'' icon on each task card \autoref{fig:task_card_page1}, users can display all corresponding generations on the canvas. Users can also click on the task name label on each task header \autoref{fig:task_status} to display generations corresponding to the specific node locally.

For a reactive microtask, generations are only requested once users click on the label on a task header. On such occasions, the label turns black once generations are ready, and users can click on the label again to hide all generations. For a proactive microtask, besides clicking on the ``expand'' icon on the ``curtain'' (\autoref{fig:notifications_and_previews}), users can also click on the label on task headers to see all resulting diagrams of a microtask. In this case, the microtask on this node is in ``display'' mode, and the text on the label is underlined. Should there be unread results, there would also be an ``expand all'' icon that users can click on to show the results of all microtasks. Should there be any microtask in ``display'' mode, there would also be a ``close all'' icon for users to hide the results of all microtasks.

\subsubsection{Toggling Initiative Modes} \label{task_initiative}
The initiative modes (proactive vs. reactive) can also be toggled both globally and locally for our \textit{Goal2} and \textit{Goal 3}. Users can either click on the task name label on the ``task card'' to deactivate it globally (\autoref{fig:task_card_inactive}), or control-click labels on each task header to deactivate the microtask for a specific node (\autoref{fig:task_status}). Once a microtask is globally deactivated, the corresponding label will turn gray and function solely as a button on all task headers, provided that it is not in ``display'' mode and has no notifications. Task headers of newly created diagrams will also have this microtask turned off.

\subsubsection{Task Configuration and Delegation}
To support our \textit{Goal 2}, we allow users to specify microtask requirements, including input and output types, and prompts.
We also enable users to rapidly create and delegate a new microtask to the LLM.

\paragraph{\underline{Configuring Task Specifications}} \label{task_configuration}
Users can change the input and output types of a microtask by clicking on the left and right arrow triangle icon (\autoref{fig:task_card}a). We currently support four input types (\textbf{\textcolor{keyword}{\textit{keyword}}}, \textbf{\textcolor{concept}{\textit{concept}}}, \textbf{\textcolor{sticky_note}{\textit{sticky note}}}, \textbf{\textcolor{section}{\textit{section}}}) and three output types (\textbf{\textcolor{keyword}{\textit{keyword}}}, \textbf{\textcolor{concept}{\textit{concept}}}, \textbf{\textcolor{sticky_note}{\textit{sticky note}}}). Users can also change prompts by double-clicking the text box of the prompt (\autoref{fig:task_card}b). For some microtasks, they can also switch between predefined prompt examples. Each prompt should have a ``[placeholder]'' to be filled with the text of a node (\autoref{fig:task_card}b).

\begin{figure*}[ht]
\includegraphics[width=\textwidth]{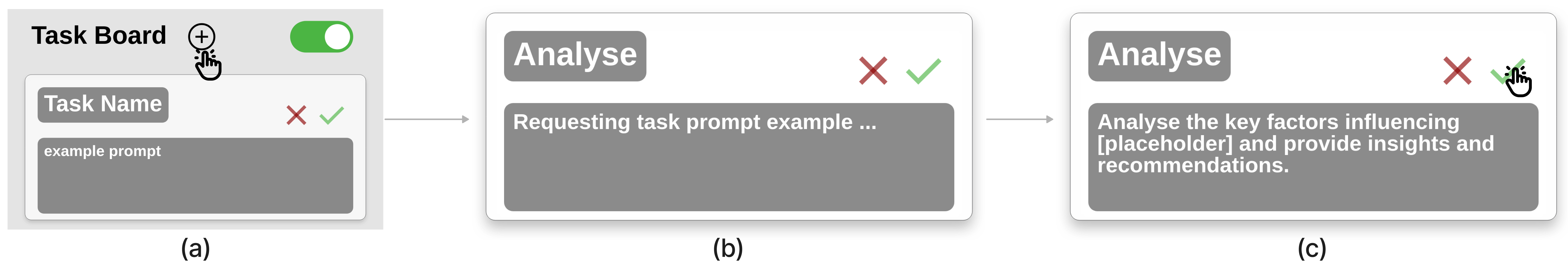}
\caption{The workflow of delegating a new microtask. (a) The user first clicks on the ``add'' icon and a new card will appear. (b) After user inputs a task name, the system requests an example prompt from the LLM. (c) User clicks ``confirm'' to add the microtask.}
\label{fig:add_new_task}
\end{figure*}

\paragraph{\underline{Delegating a New Microtask}}
For our \textit{Goal 2.3}, a user can click the ``add'' icon on the task board to create a new microtask (\autoref{fig:teaser}B). After clicking, the task board will spawn a new card for users to specify the task name and the example prompt. Users can later change input and output type, or toggle visibility or initiative on a newly added task card after clicking ``confirm''. When a user clicks to specify a task name, the system will prompt the LLM to suggest a task name. Once a task name is specified, the system will also request an example prompt for users from the LLM. \autoref{fig:add_new_task} illustrates the whole process of adding a new microtask.


\subsection{Implementation}
\textit{Polymind} was implemented using Javascript, React \footnote{https://react.dev/} and react-konva \footnote{https://konvajs.org/docs/react/Intro.html}. The backend LLM was ChatGPT (GPT-4), and the temperature was set to $0.7$.
We set $x=5$ based on our pilot study, i.e., every 5 seconds, we sample a diagram node for each of the 5 input types to be processed by proactive microtasks. If \textit{x} were too small, users would be overwhelmed with notifications of new results; if too large, there would be noticeable latency, and users will often end up waiting for results. Once the user confirms an update of a diagram node's text content, a re-sampling of the corresponding input type will be executed immediately.

To reduce the computational cost, we quit prompting ChatGPT with the sampled diagram node if a proactive microtask has unchecked results (i.e., notifications) or expanded diagrams on it.
To make generations less random, in each prompt we added constraints on its output.
\revision{For \textbf{\textcolor{keyword}{\textit{keyword}}}, we request 3 generations, each no more than 3 words; for \textbf{\textcolor{concept}{\textit{concept}}} we request 3 generations no more than 5 words; for \textbf{\textcolor{sticky_note}{\textit{sticky note}}} we request 1 generation no more than 150 words.}
Whenever ChatGPT returns results, we preserve the previous dialogue in each resulting diagram node. Once users request an explanation or a regeneration of the node, we further prompt the ChatGPT based on the saved dialogue (see \ref{appendix_feedback}). When the system requests a microtask name, we use names of predefined microtask as a few-shot template to prompt the ChatGPT. When the system requests a prompt for a new microtask, we use name-prompt pairs of predefined microtasks as a few-shot template. We refer our readers to \ref{appendix_newtask} for more details.
\section{Evaluation}
We conducted an exploratory evaluation of \textit{Polymind}, focusing on the usability, creativity, and usefulness of its parallel collaboration workflow.
More specifically, the study aimed to answer the following two questions:
\begin{enumerate}
    \item Is \textit{Polymind} easy to use, and useful for prewriting?
    \item How effective are \textit{Polymind}'s parallel collaboration workflow and microtasking features for supporting creativity in prewriting?
\end{enumerate}

\subsection{Participants}
We used convenience sampling to recruit 10 participants (4 male, 6 female) mainly from local universities. All participants are L2 English speaker.
\revision{Similar to our formative study, we mainly reached out to participants with creative writing experience or related majors, though not necessarily expert writers. All participants had experience using prewriting or diagramming tools such as Figma or Miro, but reported limited knowledge or experience of AI or programming.}
We refer to them as V1-10. For each participant, we offered a coupon equivalent to 50 HKD.

\subsection{Study Design}

\subsubsection{Tasks}
To evaluate our \textit{Polymind} in both divergent and convergent thinking phases, we divide a creative pre-writing task into two sessions: story ideation, and story outlining. In the story ideation, participants were required to brainstorm as many distinct storylines as possible. Each storyline only needs to be one or two sentences long that specifies main characters, events, locations, time, etc. In the story outlining session, participants were asked to pick one favourite storyline from the previous session, and draft a rough outline with as many details as possible. An outline needs to specify a clear structure (such as the classic beginning-climax-ending structure), and key events along the structure.

\subsubsection{Conditions}
The study compared our system, \textit{Polymind} to a turn-taking, conversational interface, plus \textit{Polymind}'s diagramming canvas, as a baseline using two creative writing prompts. \revision{The baseline allows users to take notes and keep track of conversational results using the canvas, but does not require nor allow users to interact with the AI via diagrams}. Specifically, two system conditions were used:
\begin{itemize}
    \item \textbf{GPT-4 \& Canvas} OpenAI ChatGPT-4 interface plus \textit{Polymind}'s diagramming canvas. All microtasking features were turned off.
    \item \textbf{\textit{Polymind:}} full version of \textit{Polymind} with six predefined default microtasks.
\end{itemize}
Two creative writing prompts were chosen:
\begin{itemize}
    \item Write a story where your character is traveling a road that has no end, either literally or metaphorically.
    \item Write a story in which a character is running away from something, literally or metaphorically.
\end{itemize}
We used a Latin square experimental design~\cite{ryan2007modern} to achieve a balanced sequence of writing prompts and system conditions.

\subsubsection{Study Procedure}
After giving consent to our study, users were invited to use two systems in turn to complete two sessions: story ideation and story outlining, given two different writing prompts. Each session lasted 12 minutes, and users were given time to transfer their prewriting results (generations, diagrams, or merely thoughts and ideas in their minds) to another document after each session concluded. Before using \textit{Polymind}, we walked the participants through all of its features and offered approximately 10 minutes for them to try out the system.

After two sessions concluded for a system condition, each participant was required to complete a survey, including a NASA Task Load Index (NASA-TLX)~\cite{hart1986nasa}, and three dimensions (2 questions each) of Creativity Support Index (CSI)~\cite{cherry2014quantifying}: Enjoyment, Exploration, and Expressiveness. Two dimensions (Collaboration \& Immersion) of the original CSI were dropped because they were irrelevant to the two questions we sought to answer.

To evaluate the final results (outlines), we invited two expert writers to score participants outlines using Torrance Test of Creative Writing (TTCW)~\cite{chakrabarty2023art}, instead of the self-rated score of CSI. \revision{We did a quick interview after the scoring to ask about their general feedback, and to compare results of two conditions and pick out examples with most noticeable differences.}
One of our expert is a professional fiction writer and has a doctoral degree in film studies. She used to be a screenwriter before becoming a fiction writer. The other expert is an AO3 (Archive of Our Own) writer that has posted over 400K words and accumulated over 250K views.

After using two system conditions (4 sessions in total), each participant was then required to complete a survey to rate the usefulness of each \textit{Polymind} features. We then conducted a brief interview (5-10 min) to ask about their overall use experience, feedback on \textit{Polymind}'s workflow, perceptions of creativity support, and perceived differences between the two workflows and their impact on the final results.

The whole study procedure lasted around 2 hours, and was screen recorded. The interviews were audio-taped and transcribed for analysis.

\subsection{Study Results}
In this subsection, we report the findings of our study. On balance, \textit{Polymind}'s parallel microtasking workflow granted more customizability and was more controllable. Therefore users reported a stronger sense of agency, ownership of results, and a higher level of expressiveness. The microtasking workflow could also help quickly expand idea trees through ``chaining''-like effects~\cite{wu2022ai}.


\begin{figure*}[htb]
\centering
\includegraphics[width=.64\linewidth]{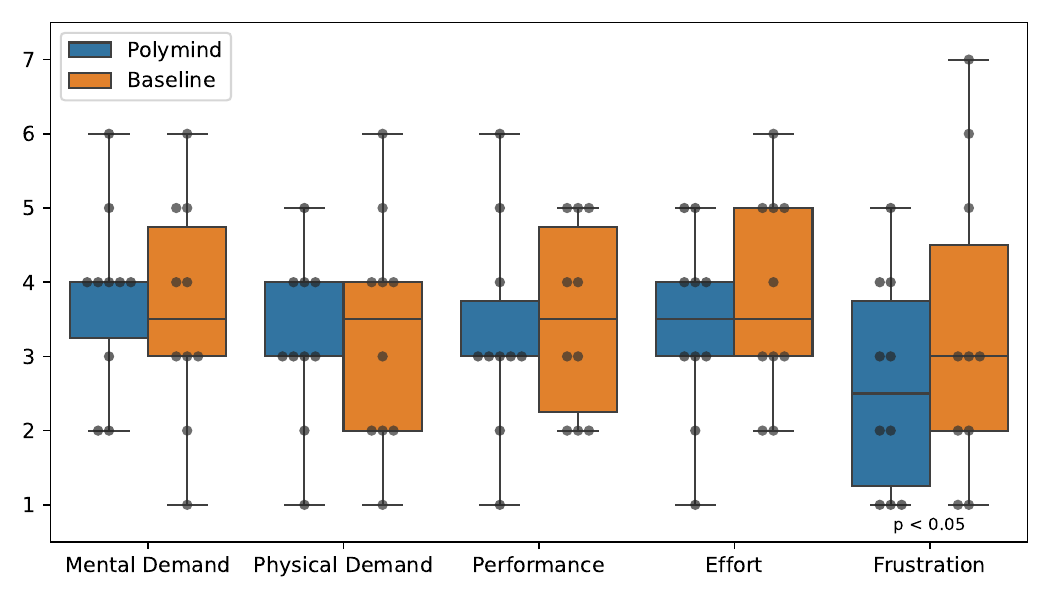}
\caption{NASA Task Load Index of \textit{Polymind} and Baseline conditions (the lower, the better).}
\label{fig:usability}
\end{figure*}

\subsubsection{Usability \& Usefulness}
Despite efforts of microtask and diagram management and the potential learning curve, to our surprise, \textit{Polymind} was almost perceived as easy to use as the ChatGPT interface, and significantly reduced frustration towards generated results (as shown \autoref{fig:usability}). ChatGPT interface was demanding mainly due to efforts of digesting longer text information (e.g., V4-5), and typing and iteratively refining lengthy prompts (e.g., V1-2). \revision{For \textit{Polymind}, the main cause of demand was said by V2-3, \& V10 to be the efforts of mannually managing the canvas, adjusting its layout, and progressing through diagrams.}
Notably, V1 \& V2 said that \textit{Polymind}'s interface was easier to navigate, and easier to read, because its generations were mainly short phrases, and had structures (including lines, sections, \& microtask colours). 
Besides, the randomness of ChatGPT generations also caused higher frustration level and worse perceived performance than \textit{Polymind} among some participants (e.g., V5, V8).

\begin{figure*}[htb]
\centering
\includegraphics[width=.85\linewidth]{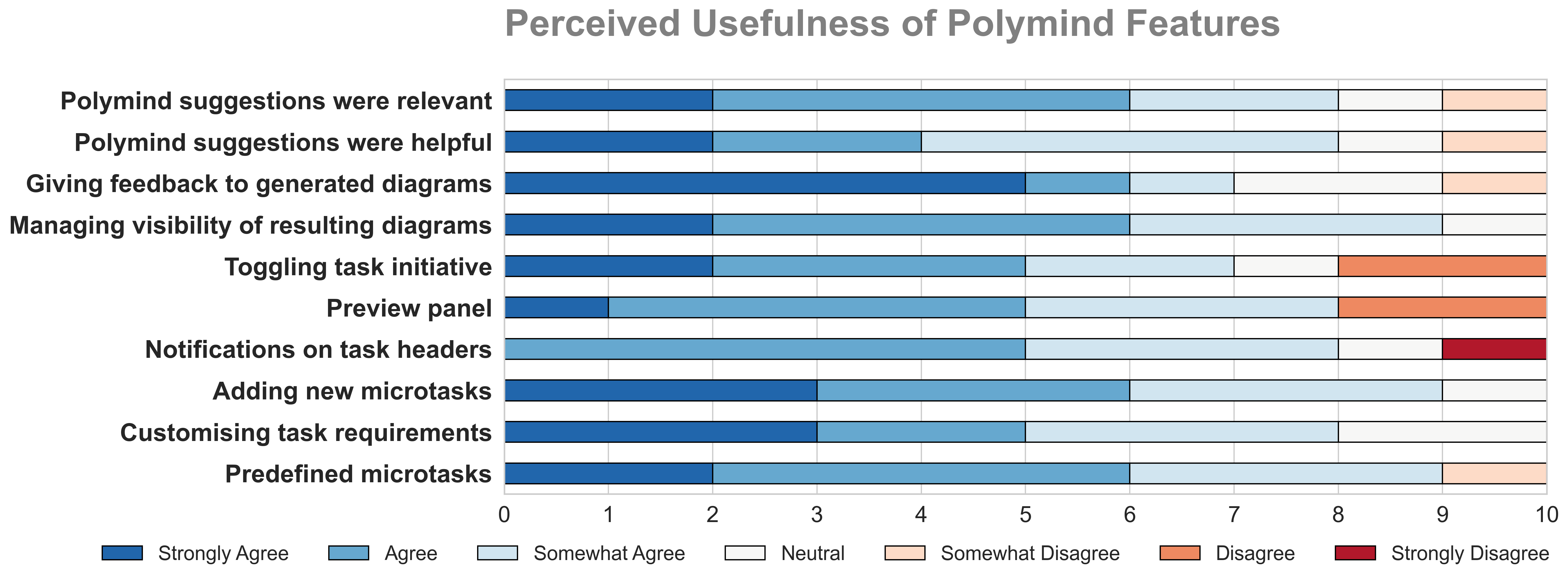}
\caption{The perceived usefulness of \textit{Polymind} features}
\label{fig:usefulness}
\end{figure*}

The key features of \textit{Polymind} were mainly perceived useful for prewriting, as shown in \autoref{fig:usefulness}. Of them awareness-related features, such as notifications, previews, and initiative modes, were found most controversial, which revealed the tension of our \textit{Goal 2.1} and being overall non-intrusive. V3 felt a proactive microtask was particularly annoying and intrusive, but we observed that all other participants left key microtasks proactive. Some said (e.g., V5, V7) they would need proactive microtasks in divergent thinking phases for quick ideas, but sometimes did not want to be interrupted while thinking. Therefore, most participants thought the feature of switching intiative modes particularly helpful.

\subsubsection{Creativity Support}
\begin{figure}[ht!]
\centering
\begin{minipage}[t]{.4745\linewidth}
    \vspace{0pt}
    \includegraphics[width=\linewidth]{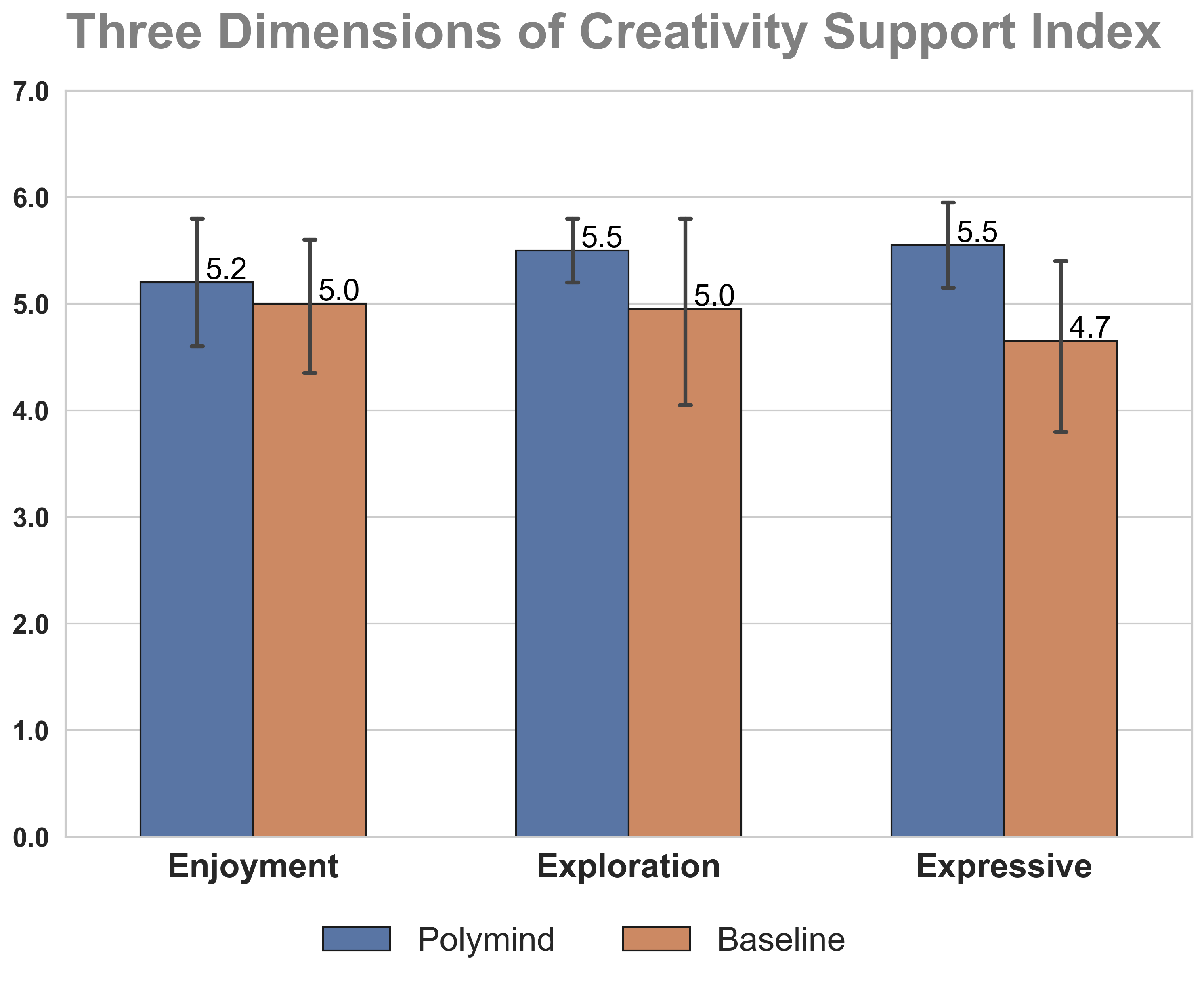}
\end{minipage}
\begin{minipage}[t]{.32\linewidth}
    \vspace{0pt}
    \includegraphics[width=\linewidth]{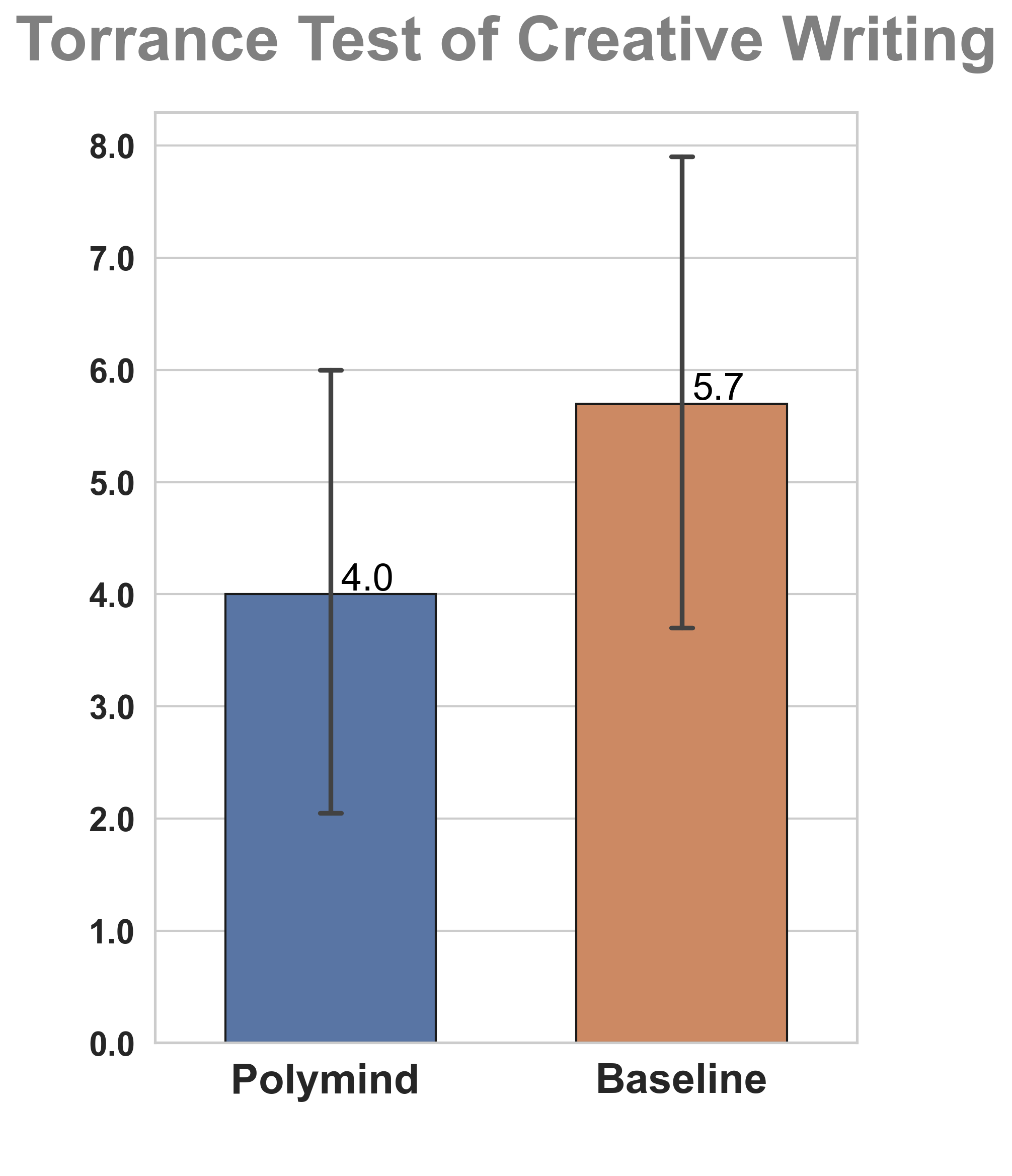}
\end{minipage}
\caption{The results of Creativity Support Index (CSI) and Torrance Test of Creative Writing (TTCW)}
\label{fig:CSI}
\end{figure}

In terms of creativity, participants generally felt \textit{Polymind} was more supportive (see \autoref{fig:CSI}), but the results were not significant ($P_{Enjoyment}=0.67$, $P_{Exploration}=0.19$, $P_{Expressiveness}=0.05$). Notably, the expressiveness dimension has almost shown significance, as many (e.g., V3 \& V5) reported that \textit{Polymind}'s diagramming interface and microtasking workflow put them in dominant roles that encouraged them to freely express their brief ideas. In terms of results, two conditions produced similar number of ideas ($Polymind_{median}=3$, $Polymind_{stdev}=1.06$, $Baseline_{median}=3$, $Baseline_{stdev}=8.51$) in the ideation session.

In addition, experts' scores showed that the baseline condition produced outlines that were able to pass 5.7 TTCW tests, as compared to \textit{Polymind}'s 4 tests ($P=0.23$) (see \autoref{fig:CSI}).
\revision{
Experts did not particularly mention any noticeable differences in quality between two conditions except that \textit{Polymind}'s results were much shorter and lacked details to pass some tests. Besides, they both expressed concern of overused or clichéd results. One expert said she was initially interested by V2's story (baseline), but only to find out that it was from \textit{The Vampire Diaries}
}
This is expected, as ChatGPT interface could quickly generate ``\textit{complete and detailed outlines with simple prompts}'' (V2), while \textit{Polymind} usually encouraged users to make progress in diagrams with limited words.
Although some (e.g., V8) noted that they could still generate a complete outline using \textit{Polymind}, but they simply did not want to, because they would like to take control, and create a story from their own fragmented ideas, instead of borrowing all results from ChatGPT.

\subsubsection{Microtask Usage: Quick Chaining and Idea Expansion}
\revision{All default microtasks have been applied by 10 participants to produce their final results, as shown in \autoref{tab:usage}. In some cases users might have default microtasks slightly edited. For example, V2 changed the prompt of \BboxS{\textcolor{white}{Brainstorm}} and switched the output type to \textbf{\textcolor{sticky_note}{\textit{sticky note}}} to request detailed settings of a story. Four participants have delegated a total of 8 customized microtasks. For example, V8 delegated \CboxS{\textcolor{white}{Beginning}} \& \CCboxS{\textcolor{white}{Climax}} to generate a beginning and climax of a given storyline. V4 delegated \CboxS{\textcolor{white}{Juice}} to juice up a given story in a \textbf{\textcolor{sticky_note}{\textit{sticky note}}} with more details.}

\revision{We also found participants came up with creative and efficient ways of using a combination of microtasks}. By leaving some microtasks in the proactive mode, \textit{Polymind} can easily perform the ``chaining'' operation~\cite{wu2022ai} to expand users' ideas in a tree-like structure. During this process, multiple distinct microtasks could contribute simultaneously in parallel to users' main operations, which made the collaboration more efficient and creative.
Some participants complimented that the parallel microtasks were like ``\textit{a mature pipeline that needs little efforts}'' (V5), ``\textit{as if splitting (brainstorming) indefinitely}'' (V6). For example, V2 mainly used two proactive microtasks: \FboxS{\textcolor{white}{Freewrite}} \& \SboxS{\textcolor{white}{Summarise}} during the story ideation session.
\revision{He later explained that \FboxS{\textcolor{white}{Freewrite}} was used to quickly generate stories in a \textbf{\textcolor{sticky_note}{\textit{sticky note}}} given a few keywords or concepts within a \textbf{\textcolor{section}{\textit{section}}}, while \SboxS{\textcolor{white}{Summarise}} presented brief summaries in a \textbf{\textcolor{sticky_note}{\textit{sticky note}}} of \FboxS{\textcolor{white}{Freewrite}}'s generations so that he would not need to read whole stories.}

V10 instead was mainly using \BboxS{\textcolor{white}{Brainstorm}} and \EboxS{\textcolor{white}{Elaborate}} to expand her ideas in brief keywords and concepts. \revision{She used \BboxS{\textcolor{white}{Brainstorm}} to request related ideas and \EboxS{\textcolor{white}{Elaborate}} to provide concrete examples of an idea.} In a comparison to the ChatGPT interface, she commented that,
\begin{quote}
    ``\textit{I feel that ChatGPT often generated something irrelevant, and missed my expectations. But this system (Polymind) stuck to my main concept by generating relevant ideas. Although the results were only brief keywords, but it was fast. It could produce a huge idea tree within a short period of time, and you could easily find something intriguing and figure out a coherent story.}''.
\end{quote}

\newtcbox{\Bbox}{on line,
  colframe=brainstorm,colback=brainstorm,
  boxrule=0.5pt,arc=1pt,boxsep=0pt,left=2pt,right=2pt,top=2pt,bottom=2pt}
\newtcbox{\Sbox}{on line,
  colframe=summarise,colback=summarise,
  boxrule=0.5pt,arc=1pt,boxsep=0pt,left=2pt,right=2pt,top=2pt,bottom=2pt}
\newtcbox{\Ebox}{on line,
  colframe=elaborate,colback=elaborate,
  boxrule=0.5pt,arc=1pt,boxsep=0pt,left=2pt,right=2pt,top=2pt,bottom=2pt}
\newtcbox{\Dbox}{on line,
  colframe=draft,colback=draft,
  boxrule=0.5pt,arc=1pt,boxsep=0pt,left=2pt,right=2pt,top=2pt,bottom=2pt}
\newtcbox{\Fbox}{on line,
  colframe=freewrite,colback=freewrite,
  boxrule=0.5pt,arc=1pt,boxsep=0pt,left=2pt,right=2pt,top=2pt,bottom=2pt}
\newtcbox{\Abox}{on line,
  colframe=associate,colback=associate,
  boxrule=0.5pt,arc=1pt,boxsep=0pt,left=2pt,right=2pt,top=2pt,bottom=2pt}
\newtcbox{\Cbox}{on line,
  colframe=custom,colback=custom,
  boxrule=0.5pt,arc=1pt,boxsep=0pt,left=2pt,right=2pt,top=2pt,bottom=2pt}
\newtcbox{\CCbox}{on line,
  colframe=custom2,colback=custom2,
  boxrule=0.5pt,arc=1pt,boxsep=0pt,left=2pt,right=2pt,top=2pt,bottom=2pt}

\renewcommand{\arraystretch}{1.2}
\begin{table*}[htb]
    \resizebox{\linewidth}{!}{
    \begin{tabular}{c|cc|cc}
    \toprule
    & \multicolumn{2}{c|}{Task \RNum{1}} & \multicolumn{2}{c}{Task \RNum{2}} \\
    & default & custom & default & custom \\
    \midrule
    V1 & \Ebox{\textcolor{white}{Elaborate}} \Fbox{\textcolor{white}{Freewrite}} & \Cbox{\textcolor{white}{Characteristics}} & \Sbox{\textcolor{white}{Summarise}} \Fbox{\textcolor{white}{Freewrite}} & \\
    V2 & \Bbox{\textcolor{white}{Brainstorm}} \Sbox{\textcolor{white}{Summarise}} \Fbox{\textcolor{white}{Freewrite}} & & \Ebox{\textcolor{white}{Elaborate}} \Fbox{\textcolor{white}{Freewrite}} & \Cbox{\textcolor{white}{Structure}} \\
    V3 & \Bbox{\textcolor{white}{Brainstorm}} \Fbox{\textcolor{white}{Freewrite}} \Dbox{\textcolor{white}{Draft}} \Abox{\textcolor{white}{Associate}} & & \Bbox{\textcolor{white}{Brainstorm}} \Sbox{\textcolor{white}{Summarise}} \Fbox{\textcolor{white}{Freewrite}} & \\
    V4 & \Bbox{\textcolor{white}{Brainstorm}} & \Cbox{\textcolor{white}{Structure}} \CCbox{\textcolor{white}{Juice}} (up) & \Bbox{\textcolor{white}{Brainstorm}} & \Cbox{\textcolor{white}{Structure}} \CCbox{\textcolor{white}{Theme}} \\
    V5 & \Bbox{\textcolor{white}{Brainstorm}} \Fbox{\textcolor{white}{Freewrite}} \Abox{\textcolor{white}{Associate}} & & \Bbox{\textcolor{white}{Brainstorm}} \Fbox{\textcolor{white}{Freewrite}} \Abox{\textcolor{white}{Associate}} & \\
    V6 & \Bbox{\textcolor{white}{Brainstorm}} \Abox{\textcolor{white}{Associate}} & & \Bbox{\textcolor{white}{Brainstorm}} \Abox{\textcolor{white}{Associate}} & \\
    V7 & \Bbox{\textcolor{white}{Brainstorm}} & & \Bbox{\textcolor{white}{Brainstorm}} \Ebox{\textcolor{white}{Elaborate}} & \\
    V8 & \Bbox{\textcolor{white}{Brainstorm}} \Dbox{\textcolor{white}{Draft}} \Abox{\textcolor{white}{Associate}} & & \Bbox{\textcolor{white}{Brainstorm}} \Dbox{\textcolor{white}{Draft}} \Fbox{\textcolor{white}{Freewrite}} & \Cbox{\textcolor{white}{Beginning}} \CCbox{\textcolor{white}{Climax}} \\
    V9 & \Bbox{\textcolor{white}{Brainstorm}} \Dbox{\textcolor{white}{Draft}} \Abox{\textcolor{white}{Associate}} & & \Dbox{\textcolor{white}{Draft}} & \\
    V10 & \Bbox{\textcolor{white}{Brainstorm}} \Ebox{\textcolor{white}{Elaborate}} \Abox{\textcolor{white}{Associate}} & & \Bbox{\textcolor{white}{Brainstorm}} \Abox{\textcolor{white}{Associate}} & \\
    \bottomrule
    \end{tabular}}
    \caption{Microtasks used by each participant in \textit{Polymind} condition during two sessions that produced the final results. The input \& output types and prompts of default microtasks might have been edited.}
    \label{tab:usage}
\end{table*}

V3 was one of the three participants that showed clear preference for the ChatGPT interface, but she also added that experimenting ideas with \textit{Polymind} were much easier. She explained that \textit{Polymind} ``\textit{had a structure}'' and could perform chaining-like operations easily ``\textit{by using sections}'' and parallel microtasks, while for ChatGPT, ``\textit{combining elements (like some characters, events, or settings) from its generations to re-prompt it was challenging}''. Similarly, V6 noted that brainstorming associations between key events or scenes was much easier with \textit{Polymind} by ``\textit{using several proactive microtasks operating on nodes or sections}''.

\subsubsection{Polymind is More Controllable}
While ChatGPT-4's conversational interface was able to generate long pieces of text with many ideas and details (V2-4, V7-9), most of our participants (V1-2, V4-5, V7-8, V10) mentioned that \textit{Polymind} felt more controllable in a prewriting task. This is because \textit{Polymind} directly operated on diagrams that were often shorter and \revision{thus easier to digest and re-prompt} than ChatGPT-4's conversations, and the canvas progressed in a structured manner with parallel microtasks handling very specific requirements.

The longer generations of the conversational interface were often criticised for being hallucinatory (e.g., ``\textit{not that creative or sensible as it appears}'' -- V7), too random (e.g., ``\textit{not what I expected}'' -- V5, ``\textit{irrelevant}'' -- V10), or ``\textit{mediocre}'' (V2) during a story pre-writing session. In comparison, \textit{Polymind} generations were perceived by many to be relevant (V2, V5, V10), and its microtasks more responsive to users' requests (V5, V8, V10), although it might require some efforts to manage or configure them (V4). \revision{This echoes with the usability score of the two conditions, where \textit{Polymind} was on the same level with the baseline, despite the efforts of managing a diagramming interface.}
V8 said in retrospect that,
\begin{quote}
    ``\textit{I think this (Polymind) would be very helpful for coming up with a story. Cause you can specify your beginning, you can specify your climax, and the ending too... And I think customizing microtasks is also a nice feature... You can really outline everything. While for GPT, you often don't know what is beginning, what is climax or ending.}''
\end{quote}

Additionally, V1, V2 and V8 noted that \textit{Polymind}'s microtasking workflow made it easier to re-prompt. V8 said,
\begin{quote}
    ``\textit{If I want to change something, I know where the part is. Like the character, I only need to change several keywords, like, Oh I'd like the character to be a dragon... I felt that my prompts were actually considered (by microtasks), while GPT sometimes doesn't process all my prompts.}''
\end{quote}
For the conversational interface, it often took multiple iterations to reach a decent draft (V4-5), and each prompt had to be lengthy to change the context (V1-2, V5), which was demanding. V2 also added that the \textit{Polymind} interface was neater because with some parallel microtasks it required little to no efforts of note taking to ask multiple follow-up questions of different ideas.

It is worth noting that, three participants that disliked \textit{Polymind}'s workflows mentioned that it was quite demanding sometimes to configure microtasks (V3-4), and progress in diagrams (V9). V9 said, ``\textit{GPT could generate a lot with a single prompt, while \textit{Polymind} only little by little.}''
\revision{V10 shared similar sentiments. She noted prompting ChatGPT would be much easier than using \textit{Polymind}'s diagrammatic workflow if its generations were not random. However, she added that \textit{Polymind} was in reality less demanding because you could explore more options and easily drop random results.}

\subsubsection{Polymind Affords Agency}
Our participants almost unanimously said that \textit{Polymind} put users in a dominant role, while with the conversational interface they were completely guided by the GPT. This aligns with our \textit{Goal 2} that aims to put humans in a role of managing all microtasks. Participants without any ideas, such as V3 \& V4, generally did not mind following the ChatGPT. This is expected, and agrees with our formative study. However, same as almost all other participants, they particularly mentioned that they felt these results were not their ideas, as ChatGPT generated almost everything. While using \textit{Polymind}, participants said that they had more freedom and control (V3, V6-8), and needed to think a lot (V4-5, V8). 

One of the participants, V5, particularly said he had no trust in AI because ``\textit{it could not be truly creative}''. He therefore became very annoyed with the ChatGPT interface when it did not generate what he expected, saying it was ``\textit{bad usability}'', while attributing ``\textit{good usability}'' to the task management workflow. V7 also expressed concerns for using the conversational interface for brainstorming,
\begin{quote}
    ``\textit{At first sight, it might seem it had generated everything you could think of, but then you'd find many were indeed non-sensical. But I felt I was confined to these generations after reading them. It was especially hard for a novice writer like me to come up with other possibilities. So it felt like it was GPT that was composing a fiction, rather than me.}''
\end{quote}
She later added that her own results from \textit{Polymind} felt more ``\textit{logical}'', and ``\textit{rigorous}''.

Of the participants that said very positively of the ChatGPT interface, V3 stressed that \textit{Polymind} encouraged her to express her own ideas, while ChatGPT did not. That was why she assigned a very low score of expressiveness in the CSI survey when using ChatGPT.
\section{Discussion}
In this section, we discuss insights of our design and study, and reflect on potential future work.
\subsection{Revisiting AI Initiative}
Previous literature suggests that AI can be frustrating and cause breakdowns and distrust if it fails to understand users' intentions while taking the initiative \cite{buschek2018researchime,clark2018creative,oh2018lead}. Our evaluation further reveals that the preferences of initiative modes are most likely task-dependent.
Considering these challenges, we suggest that a generative AI should be more transparent about its task-specific limitations \cite{smith2022real}, and dynamically adjust its level of initiative accordingly. A generative AI could proactively provide suggestions on tasks it is good at, but it should play mostly a supporting role on tasks where it might be unreliable.

Furthermore, we believe that initiative modes of future generative AI design should go beyond a simple on/off switch. Task-specific capabilities of AI can vary greatly across a continuous spectrum; therefore, the control of AI initiative should also be at least partly continuous. For example, while an AI is taking the initiative and proactively providing suggestions, it can signal uncertainty and make suggestions more subtle and non-intrusive when the user switches to a task it is not familiar with or good at, instead of being completely ``turned off''.
We might also consider designing features to support users in managing initiative levels of AI instead of fully automating it.

Additionally, while supporting users through microtasking, or with multiple functions or roles, the design of AI initiative should also take into account the frequency of use (as suggested by S6), the intrusiveness of potential generations (as suggested by S8), and users' possible processing order of microtasks (as suggested by S10), based on the nature of specific tasks.
Both our formative study and system evaluation also suggest that users are more likely to let AI take control in divergent thinking tasks such as brainstorming. Therefore, it might be reasonable to apply different initiative modes for divergent and convergent thinking scenarios: AI-dominant mode for divergent thinking, and user-dominant mode for convergent thinking. The corresponding microtask management features should also be adapted in accordance with the human-AI dynamics.

\subsection{Designing Awareness Features} \label{designing-awareness-features}
Our study unveils the challenge of balancing awareness features of AI (or multiple AI functions or roles) and the overall non-intrusiveness of the collaboration workflow. This is especially challenging on a diagramming canvas because an operation might occur on an arbitrary node, thus calling for users' attention from arbitrary locations. In general, our participants perceived our awareness features to be non-intrusive, but it is worth pointing out that this perception might be biased due to participants finding microtask suggestions to be overall helpful and relevant. Our findings also suggest that users might not always be satisfied with a minimal preview or notification as awareness information:
some found it could be hard to notice, but enlarging it might be distracting.

To this end, we believe a future study is needed to investigate the specific design requirements of awareness features for generative AI results on other diagram-based (or even canvas-based in general) interface beyond prewriting (e.g., sketch-based diagramming \cite{zeleznik2008lineogrammer,kara2004hierarchical}, visual programming \cite{li2020supporting}, etc.). It is also worth studying how attention and interruption should be balanced on such interfaces~\cite{gluck2007matching}. Furthermore, we believe awareness features should go beyond animations of discrete levels of attention-drawing strength. It is reasonable to assume that suggestions with higher quality deserve more attention.
\revision{Our user study suggests that perceptions of result quality can have a significant impact on perceived mental demand. For creative tasks such as prewriting, ``quality'' implies both originality and relevance. How to take into account these contextual factors in awareness design poses yet another significant question.}

\subsection{Diagramming and Prewriting}
Our evaluation shows that a microtasking workflow on a diagramming canvas can efficiently provide creative writing ideas in an organised and non-interrupting manner. This aligns with previous results about parallel input workflows in human collaboration scenarios, which found them to be more creative than serial interactions~\cite{shih2009groupmind,hymes1992unblocking}. Microtasking complements Jiang et al.~\cite{jiang2023graphologue}'s diagrammatic workflow by facilitating parallel collaboration to enhance creativity. The implementation of this workflow, including awareness features, initiative modes, etc., could set a paradigm for future endeavor to incorporate LLMs into diagramming interfaces for prewriting.

Furthermore, the choices of predefined microtasks in \textit{Polymind}'s workflow represented a range of prewriting needs, which covered both convergent and divergent modes of thinking in a creativity process. Findings of our study suggests that, \textit{Polymind} could not only provide more ideas, but also effectively organise or synthesize existing information on the canvas for further iterations. Users were able to quickly generate a first draft by sectioning over a set of diagrams, or request new ideas based on a summary of existing diagrams. This finding might inspire future design to transform prewriting workflows, including but not limited to visual diagramming. Although prewriting by nature is iterative not linear~\cite{rohman1965pre}, most previous prewriting tools divides the prewriting process into distinct linear phases: Lu et al.'s Inkplanner roughly has three phases: diagramming, structuring, and outlining~\cite{lu2018inkplanner}; Sadauskas's design covers three phases: content aggregation, content re-experience, and narrative development~\cite{sadauskas2015mining}. As a result, these tools might not accommodate intermediate results well, especially those generated in iterations while collaborating with LLMs. Additionally, as suggested by previous studies of diagrammatic prewriting~\cite{davies2011concept,lorenz2009using}, efficient information synthesis is key to fruitful results. Future design can further consider to support organising and digesting information of complex diagram information in an iterative prewriting workflow.

\subsection{Generalizability and Future Work}
Although the design of \textit{Polymind} was motivated by parallel thinking strategies for creativity and prewriting, we expect that the microtasking workflow proposed can generalise to other diagramming interface, or other scenarios beyond prewriting or writing to enable parallel collaboration with generative AI models. \revision{As our evaluation suggests, the microtasking workflow affords multiple small, distinct, and manageable interactions at the same time for enhanced efficiency and creativity, but requires significant efforts of configuring, and compromises in-depth serial interactions.}
To apply such workflows to other human-AI collaboration tasks, the task management features with respect to our \textit{Goal 2} need to be adapted to the new context, which entails the redesign of input, output, and prompts specification, and features to support progress tracking, feedback and awareness.

For a writing task, for example, we anticipate that a microtasking interface will enhance the overall writing efficiency, quality, and creativity. Future work can adapt NLP models to accommodate writing tasks at differnt structural levels, such as sentence-level proofreading, paragraph-level summary, document-level ideation, etc. Task management features can be designed based on traditional group writing scenarios (e.g., Google Docs) \cite{birnholtz2012tracking,birnholtz2013write}. Users can delegate their own customized microtasks, or configure task specifications, including input or output types, initiative modes, prompts, etc.

For other scenarios such as drawing or design prototyping, we need to take into account the affordances of AI models across different modalities, as well as the nature of the task or interface.
For instance, on a drawing canvas, we can use multiple AI models for different purposes, such as sketch completion \cite{shi2020emog}, colorization \cite{yan2022flatmagic}, style transfer \cite{wu2023styleme}, etc. Each AI model might have unique input and output types, so the progress visualization, initiative modes, and awareness features, such as previews or notifications, should be properly redesigned on the drawing canvas.

\subsection{Limitations}
Despite promising results of our user study, our system has several limitations that should be taken into account.
First, the predefined microtasks we utilized were primarily based on existing literature on writing and creativity and may not cover all possible real-life prewriting scenarios, although users can define their own microtasks as needed. Second, our user evaluation was conducted as a controlled session of around 1 hour, which may not fully reflect the challenges and benefits of using our system in actual prewriting settings. 
\revision{Third, due to time constraints and the complexity of our evaluation task, the structure of the resulting diagrams was simple. 
While our participants did not report cognitive overload, future work could investigate users' cognitive load in more involved prewriting scenarios that would produce more complex diagrams.
To deploy our prototype in real-life scenarios, we should also consider using auto-layout algorithms to help semi-automatically arrange diagrams on the canvas.
}

\section{Conclusion}
This work explored the design and evaluation of \textit{Polymind}, which leverages large language models to support visual diagramming during prewriting. We introduced the notion of ``microtasking'' into human-AI collaborative diagramming, and designed a set of features to support managing these microtasks in parallel, including ``task headers'', ``task cards'', ``notifications and previews'', etc. Our evaluation tentatively reveals the usefulness and the potential of such a system and workflow and opens up opportunities for future research endeavors.

\bibliographystyle{ACM-Reference-Format}
\bibliography{references}


\end{document}